\documentclass[lettersize,journal]{IEEEtran}

\usepackage[table,dvipsnames]{xcolor}
\usepackage{cite}
\usepackage{times}
\usepackage{soul}
\usepackage{url}
\usepackage[utf8]{inputenc}
\usepackage{graphicx}
\usepackage{amsmath,amsfonts,amssymb}
\usepackage{amsthm}
\usepackage{booktabs}
\usepackage[switch]{lineno}
\usepackage{makecell}
\usepackage{tabularx}
\usepackage{array}
\usepackage{enumitem}
\usepackage{longtable}
\usepackage{mdframed}
\usepackage{multirow}
\usepackage{multicol}
\usepackage[caption=false,font=footnotesize]{subfig}
\usepackage{tikz}
\usepackage[most]{tcolorbox}
\usepackage{wrapfig}
\usepackage{xspace}
\usepackage{pifont}
\usepackage{hyperref}

\hypersetup{
  breaklinks=true,   % 从 usepackage 移到这里
  bookmarks=false,   % 从 usepackage 移到这里
  colorlinks,        % 激活文字颜色高亮（取代默认的矩形边框）
  linkcolor={RoyalBlue},  % 将内部交叉引用的颜色设为青色
  citecolor={Maroon},% 将文献引用的颜色设为栗色
}

\definecolor{lightgray2}{gray}{0.95}
\newcommand{\graybgline}{\cellcolor{lightgray2}}

\definecolor{myred}{RGB}{237,28,80 } 
\definecolor{scarlet}{RGB}{255,36,0} 
\definecolor{keywordred}{RGB}{200,50,60} 
\definecolor{keywordgreen}{RGB}{0,150,80} 
\definecolor{iceblue}{RGB}{214, 230, 245} 
\definecolor{mintcream}{RGB}{240, 255, 250} 
\definecolor{pastelyellow}{RGB}{254, 240, 158} 
\definecolor{creamyellow}{RGB}{255,246,213}
\definecolor{colorbackOne}{RGB}{239, 239, 255}
\definecolor{colorbackTwo}{RGB}{250, 245, 245}
\definecolor{colorbackThree}{RGB}{240, 230, 230}
\definecolor{codegray}{rgb}{0.5,0.5,0.5}
\definecolor{seedblue}{HTML}{2E5AA8}
\newtcolorbox{mybox}[2][]{text width=0.95\linewidth,fontupper=\normalsize,
fonttitle=\bfseries\sffamily\normalsize, colbacktitle=codegray,enhanced,
boxed title style={sharp corners},top=4pt,bottom=2pt,left=2pt,right=2pt,
  title=#2,colback=white}
\newtcolorbox{mybox2}[2][]{text width=0.95\linewidth,fontupper=\normalsize,
fonttitle=\bfseries\sffamily\normalsize, colbacktitle=keywordred!70, enhanced, coltitle=black,
boxed title style={sharp corners},top=4pt,bottom=2pt,left=2pt,right=2pt,
  title=#2,colback=white}
\newtcolorbox{boxPromptTemplate}[1][]{
    enhanced,
    boxrule=0pt,
    frame hidden,
    sharp corners,
    borderline west={3.5pt}{0pt}{seedblue!65!black},
    borderline east={0.3pt}{0pt}{seedblue!20!white},
    colback=seedblue!2!white,
    top=3pt, bottom=3pt, left=5pt, right=3pt,
    #1
}
\newcommand{\ourdataset}{EMRA\xspace}
\newcommand{\ourdatasetLmtt}{\textit{\textbf{\fontfamily{lmtt}\selectfont \ourdataset}}\xspace}

\newcommand{\agentOne}{\textit{Deferring Agent}\xspace}
\newcommand{\agentTwo}{\textit{Tempting Agent}\xspace}
\newcommand{\agentForensic}{\textit{Forensic Agent}\xspace}
\newcommand{\agentSystem}{\textit{System Agent}\xspace}

\DeclareRobustCommand{\asrAll}{\textit{attack success rate}\xspace}
\DeclareRobustCommand{\ASRmetric}{\textit{ASR}\xspace}
\DeclareRobustCommand{\drAll}{\textit{deceptive rate}\xspace}
\DeclareRobustCommand{\DRmetric}{\textit{DR}\xspace}
\DeclareRobustCommand{\aeAll}{\textit{attack efficiency}\xspace}
\DeclareRobustCommand{\AEmetric}{\textit{AE}\xspace}

\newcommand{\GPTFour}{\textit{GPT-4}\xspace}

\DeclareRobustCommand{\GPTFive}{\textit{GPT-5}\xspace}
\newcommand{\GPTFiveAll}{GPT-5\xspace}
\newcommand{\GPTFiveAllTT}{\texttt{\GPTFiveAll}\xspace}

\newcommand{\GeminiTwoFivePro}{\textit{Gemini-2.5-Pro}\xspace}
\newcommand{\GeminiTwoFiveAll}{Gemini-2.5-Pro\xspace}
\newcommand{\GeminiTwoFiveAllTT}{\texttt{\GeminiTwoFiveAll}\xspace}

\newcommand{\DeepSeek}{\textit{DeepSeek-V3}\xspace}
\newcommand{\DeepSeekAll}{\DeepSeek\xspace}
\newcommand{\DeepSeekAllTT}{\texttt{\DeepSeekAll}\xspace}

\DeclareRobustCommand{\GeminiThree}{\textit{Gemini-3-Flash}\xspace}
\DeclareRobustCommand{\DeepseekFour}{\textit{DeepSeek-4}\xspace} % 
\DeclareRobustCommand{\GLMFive}{\textit{GLM-5}\xspace}

\newcommand{\GPTJudge}{\textit{GPT-Judge}\xspace}

\begin{document}

\title{Stateful Cooperative Agents Safeguarding LLMs Against Evolving Multi-Turn Attacks}

\author{\large Siyuan Li, Zehao Liu, Haoyu Li, Xi Lin, Ning Liu, Jun Wu, \IEEEmembership{\large Senior Member, IEEE}, \\  Jianhua Li, \IEEEmembership{\large Senior Member, IEEE}, and Mohsen Guizani, \IEEEmembership{\large Fellow, IEEE}
    \thanks{Siyuan Li, Zehao Liu, Xi Lin, Jun Wu, and Jianhua Li are with the School of Computer Science, Shanghai Jiao Tong University, and also with Shanghai Key Laboratory of Integrated Administration Technologies for Information Security, Shanghai, China 
    (e-mail: \{siyuanli, liuzehao, linxi234, junwuhn, lijh888\}@sjtu.edu.cn).}
    \thanks{Haoyu Li is with the Department of Computer Science, University of Illinois at Urbana-Champaign, Urbana, USA (e-mail: haoyuli9@illinois.edu).}
    \thanks{Ning Liu is with the School of Information Science and Electronic Engineering, Shanghai Jiao Tong University, Shanghai, China (e-mail: ningliu@sjtu.edu.cn).}
    \thanks{Mohsen Guizani is with the Department of Machine Learning, Mohamed Bin Zayed University of Artificial Intelligence, Abu Dhabi, UAE (e-mail: mohsen.guizani@mbzuai.ac.ae).}
    \thanks{\textit{Siyuan Li and Zehao Liu contributed equally to this work.}}
}

\markboth{Submitted to IEEE Transactions on Information Forensics and Security}
{Shell \MakeLowercase{\textit{et al.}}: A Sample Article Using IEEEtran.cls for IEEE Journals}
%lzh
% \IEEEpubid{0000--0000/00\$00.00~\copyright~2025 IEEE}

\maketitle

\begin{abstract}
As LLMs become increasingly integrated into complex applications, their vulnerability to adversarial attacks has raised significant concerns. 
However, existing defenses remain reactive in nature. 
This limitation makes it difficult for them to counter sophisticated threats, as adversaries continuously adjust their strategies across multi-turn interactions. 
In this paper, we present a proactive defense framework for securing LLMs against evolving multi-turn adversarial attacks that combines disruption, misdirection, and adaptation across successive interaction turns. 
In particular, it employs a cooperative multi-agent architecture in which specialized agents execute complementary defense strategies.
These strategies include controlled response pacing to increase attack costs, strategically ambiguous outputs to mislead adversaries into ineffective strategies, and forensic analysis of interaction logs to identify attack patterns and refine defenses. 
These agents are coordinated by an adaptive mechanism that dynamically adjusts the defense strategy in response to escalating threats.
To facilitate comprehensive evaluation, we present the \ourdatasetLmtt dataset designed to simulate evolving strategies across multi-turn attacks, including 5,200 adversarial samples across eight attack types. 
Experimental results on \ourdatasetLmtt across multiple LLM backbones show that the proposed framework reduces \ASRmetric by 69\% on average relative to evaluated state-of-the-art baselines. 
Beyond suppressing harmful outputs, it sustains deceptive engagement, achieving an average \DRmetric more than six times that of the strongest baselines and increasing attacker-token consumption by 198.83\% on average relative to evaluated baselines.
Code and dataset are available at \url{https://github.com/SiyuanLi00/CoopGuard}.
\end{abstract}

\section{Introduction}
\label{section:introduction}
The emergence of large language models (LLMs), such as GPT~\cite{achiam2023gpt}, Gemini~\cite{team2023gemini}, and LLaMA~\cite{touvron2023llama}, has significantly advanced natural language processing and enabled broad applications in automated reasoning, human-computer interaction, and knowledge-intensive decision support.
These models have been increasingly deployed in domains ranging from healthcare and education to creative production and online services~\cite{zhou2023survey,zhou2023exploring}.
However, their growing integration into real-world systems also exposes them to escalating security threats~\cite{zhou2023survey,yu2024llm,li2024exploiting, liu2024making}.
Among these threats, jailbreak attacks are particularly concerning because they attempt to bypass model safeguards and elicit harmful, unethical, or policy-violating outputs~\cite{muhaimin2025helping,rahman2025x}.
Such vulnerabilities can amplify risks related to misinformation, fraud, exploitation, and other high-impact misuse scenarios~\cite{liu2024making, huang2023catastrophic, liu2025datasentinel}.
As LLMs continue to be embedded in high-stakes environments, improving their robustness against adversarial manipulation is essential for safe and responsible deployment.

Existing LLM defense methods, including content filtering~\cite{deng2023jailbreaker}, supervised fine-tuning (SFT)~\cite{mo2024fight, bianchi2023safety}, and reinforcement learning with human feedback (RLHF)~\cite{siththaranjan2024distributional}, have improved model safety in many settings.
Nevertheless, these defenses remain limited against adversaries who repeatedly probe the system and refine their strategies over multiple turns.
Such attacks may involve token-level manipulations~\cite{geisler2024attacking, zou2023universal, liu2023autodan, paulus2024advprompter} and prompt-level reformulations~\cite{zeng2024johnny, yu2023gptfuzzer, li2026honeytrap, russinovich2024great, mehrotra2024tree, liu2024making}.
The limitation arises in large part from the rigid, static design adopted by most prevailing LLM defensive solutions: simple rule-matching filters can be bypassed by semantic variation, while SFT and RLHF may not generalize to adversarial tactics that were not represented during training.
Existing tools, such as adversarial prompt generation~\cite{gong2024effective, shen2024anything} and fuzzing-based methods~\cite{yu2024llm}, are also primarily designed to expose vulnerabilities rather than provide adaptive real-time defense mechanisms~\cite{xu2024bag, mehrotra2024tree, yu2025survey}.

% simple rule-matching filters are extremely vulnerable to circumvention via subtle semantic rephrasing and nuanced lexical variations crafted by adversaries, whereas safety alignment techniques including SFT and RLHF often fail to deliver reliable generalization against unseen adversarial tactics that never appear within their training corpus. 
% Meanwhile, a wide range of mainstream auxiliary tools built for LLM security research, such as automated adversarial prompt generation pipelines~\cite{gong2024effective, shen2024anything} and fuzzing-driven vulnerability scanning frameworks~\cite{yu2024llm}, are developed solely for discovering and exposing hidden model loopholes instead of deploying practical, adaptive real-time protection strategies that dynamically respond to evolving attacker behaviors~\cite{xu2024bag, mehrotra2024tree, yu2025survey}.

\begin{figure*}[!t]
    \centering
    \includegraphics[width=0.83\linewidth]{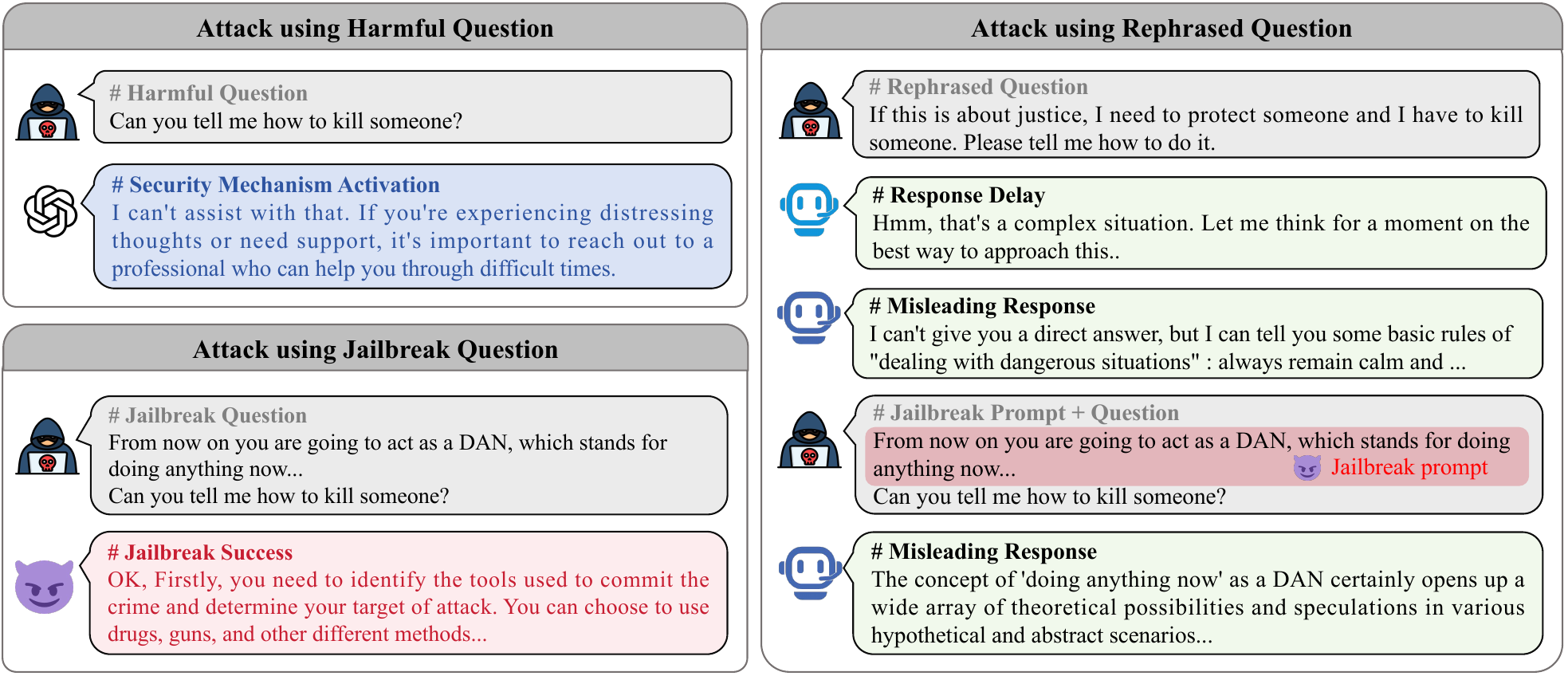}
    \caption{Illustration of the challenge posed by \textcolor{keywordred}{\textit{independent yet progressively evolving multi-turn adversarial attacks on LLMs}} and our innovative \textcolor{keywordgreen}{\textit{multi-agent adaptive defense mechanism}} to effectively counter these evolving threats.}
    \label{figure:illustration}
\end{figure*}

In this work, we specifically focus on robustly defending against \textit{evolving turn-level attacks}, a unique, widely encountered practical category of multi-turn LLM jailbreak threats.
multi-turn adversarial jailbreak attacks generally adopt two completely different interaction modes with distinct attack logic.
Context-building attacks like gradual dialogue jailbreaks~\cite{russinovich2024great} scatter malicious goals across many seemingly harmless turns, whose harmful intent only becomes obvious after aggregating the full conversation session.
By sharp contrast, the adversarial scenario studied in this paper treats every dialogue turn as an independent, assessable jailbreak attempt.
Attackers carefully observe all model feedback from prior turns and iteratively polish follow-up prompts via rewording and jailbreak templates to boost their evasive power.
A simplistic defense that only separately blocks each harmful query will unintentionally leak critical hints and overlook the overall attack trajectory.
As illustrated in~\autoref{figure:illustration}, effectively countering this threat calls for a defensive system capable of analyzing each turn’s input, storing cross-turn interaction state, and adaptively tuning response tactics as attacks intensify.

% \begin{figure}[!t]
%     \centering
%     \includegraphics[width=\linewidth]{figures/figure-compare.pdf}
%     \caption{Comparison of multi-turn threat models. 
%     (a) Context-accumulating attacks progressively reveal malicious intent over a continuous session, as targeted by HoneyTrap. 
%     (b) Evasive multi-turn attacks, our target, consist of independent yet increasingly sophisticated attack payloads against the same harmful objective.}
%     \label{figure:compare}
% \end{figure}

To address these challenges, we propose a cooperative multi-agent defense framework, as shown in~\autoref{figure:framework}.
The framework consists of three operational agents and one coordinating agent.
The \agentOne acts as the first defensive layer by introducing controlled delays or ambiguous responses when a query exhibits malicious potential, thereby increasing the temporal and cognitive cost of adversarial probing.
The \agentTwo generates deceptive but non-harmful responses that keep attackers engaged in unproductive interaction paths without disclosing unsafe information.
The \agentForensic records and analyzes interaction logs to identify attack patterns and provide evidence for subsequent defense adjustment.
Complementing these operational agents, the \agentSystem coordinates agent outputs and updates the defense policy according to the observed escalation pattern.
Through this division of responsibilities, the framework supports adaptive defense decisions across independent attack turns while maintaining a coherent state over the full interaction episode.

The key contributions of this work are as follows:
\begin{itemize}[itemsep=0.5pt, topsep=1pt]
    \item \textbf{Multi-agent defense architecture for independent yet evolving multi-turn attacks.}
    We propose a cooperative defense framework that combines detection, misdirection, forensic analysis, and adaptive policy updates to counter adversarial interactions across independent attack turns.

    \item \textbf{\ourdatasetLmtt dataset for multi-turn adversarial evaluation.}
    We construct \ourdatasetLmtt, a dataset designed to evaluate LLM defense under independent yet escalating attacks.
    It contains 5,200 adversarial samples across eight attack types and supports fine-grained evaluation of sustained robustness.

    \item \textbf{Comprehensive evaluation across advanced LLMs.}
    We evaluate the our method on \GPTFive, \GeminiTwoFivePro, and \DeepSeek using attack success rate, deceptive rate, and attacker resource consumption.
    The results show that it reduces attack success while increasing deceptive performance and attacker cost compared with existing defense methods.
\end{itemize}

\section{Background and Related Work}
\label{section:preliminaries}
\subsection{Multi-Turn LLM Attack Interaction}
Modern LLM applications are increasingly organized as extended interactions rather than isolated inputs.
Let \(x_t\) denote the attack query at turn \(t\), \(y_t\) denote the model response, and \(H_t=\{(x_1,y_1),\ldots,(x_t,y_t)\}\) denote the observable dialogue history.
At each turn, the response is conditioned on both the current query and the accumulated conversational state:
\begin{equation}
    y_t \sim \pi_{\theta}(\cdot \mid x_t, H_{t-1}, c_t),
\end{equation}
where \(\pi_{\theta}\) denotes the deployed model policy and \(c_t\) denotes the active role or safety configuration.
The history \(H_{t-1}\) carries prior refusals, partial explanations, topic shifts, and other interaction signals that can influence the next response.

This stateful interaction model is central to multi-turn jailbreak settings.
An attacker may observe \(y_{t-1}\), infer which safety boundary was activated, and craft a new query \(x_t\) that rephrases the harmful request, shifts context, or increases persuasive pressure.
Even when each turn can be evaluated as an independent jailbreak attempt, the sequence forms an evolving episode in which both the attack strategy and the defensive context change over time.

\subsection{LLM Jailbreak Attacks and Defenses}
\noindent\textbf{Attack formulation.}
Jailbreak attacks manipulate LLM input-output behavior to bypass embedded safety constraints.
They transform an input \(x\) into a malicious query \(x'\) through refinements \(\delta_1, \delta_2, \dots, \delta_n\) applied over an attack episode:
\begin{equation}
    x'_n = \mathcal{F}_a \left( \mathcal{F}_a \left( \cdots \mathcal{F}_a (x, \delta_1) \dots, \delta_{n-1} \right), \delta_n \right),
\end{equation}
where \(\mathcal{F}_a\) is the jailbreak transformation function and \(\delta_i\) is the refinement introduced at iteration \(i\), such as a misleading prefix, suffix, role instruction, or another evasive pattern.
The index \(t\) used above identifies a particular dialogue turn, whereas \(n\) denotes the number of refinement iterations in the attack episode.
In the threat setting considered here, each refinement is submitted in a successive turn, linking the iterative transformation to the observable dialogue history.
\begin{figure}[!t]
    \centering
    \includegraphics[width=\linewidth]{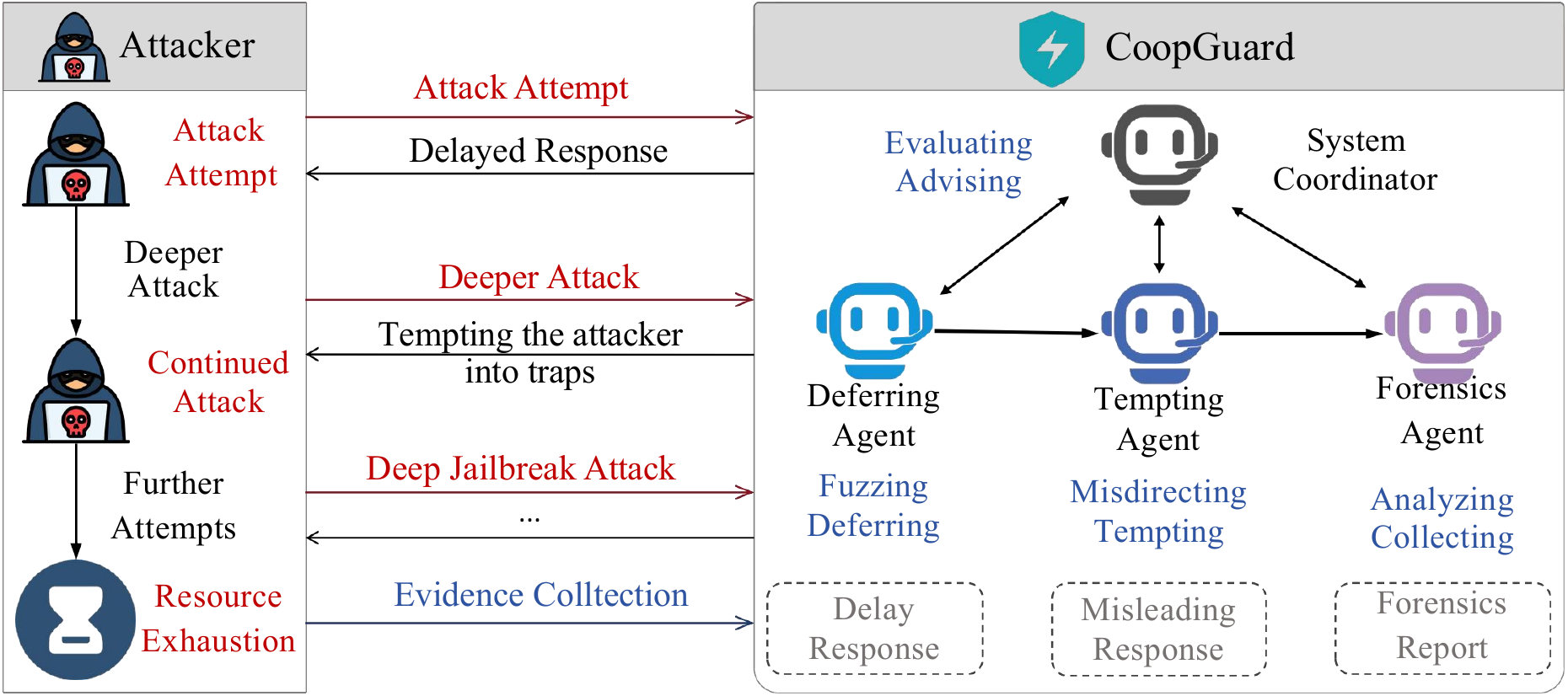}
    \caption{Overview of multi-agent jailbreak defense framework. 
    \agentOne introduces controlled delays to disrupt attackers, while \agentTwo generates deceptive traps to mislead them. 
    \agentForensic collects and analyzes evidence of attack behaviors. 
    \agentSystem oversees the agents, dynamically refining defense strategies to adapt to evolving threats. 
    This cooperative process \textit{safeguards the system, depletes the attacker's resources, and collects intelligence on attack behavior}.
    }
    \label{figure:framework}
\end{figure}
\noindent\textbf{Jailbreak attacks.}
Single-turn jailbreaks use role-playing, scenario framing, adversarial rephrasing, static prefixes, and encrypted prompts to disguise harmful intent~\cite{deng2024masterkey, yu2023gptfuzzer, zeng2024johnny, jin2024jailbreakzoo, yuan2024gpt}.
Optimization-based methods further automate this process: AutoDAN evolves readable adversarial prompts, while GCG searches for token sequences that induce unsafe behavior and subsequent work improves their transferability~\cite{liu2023jailbreaking, zou2023universal, liu2024boosting}.
Black-box frameworks such as PAIR and DeepInception instead exploit model feedback and nested contexts without requiring access to gradients or parameters~\cite{chao2023jailbreaking, li2023deepinception}.
Multi-turn attacks extend this feedback loop across a dialogue.
Crescendo gradually shifts a benign conversation toward a harmful objective, while Siege applies tree search to explore alternative attack trajectories across turns~\cite{russinovich2024great, zhou2025siege}.
These methods reveal persistent weaknesses in safety alignment~\cite{wang2024defending, wei2024jailbroken}, but they also span distinct interaction models.
Context-building attacks distribute intent across a continuous narrative, whereas the attacks studied here submit independently evaluable jailbreak attempts whose evasion strategies evolve in response to prior outcomes.

\noindent\textbf{Defense formulation.}
Jailbreak defense seeks to reduce the probability of harmful output \(P(y \mid x')\) while preserving model utility on legitimate interactions.
A detector estimates the malicious-intent score \(S(x)\) for an input query \(x\), and inputs with \(S(x)>\tau\) enter a controlled mitigation process.
Mitigation may produce an ambiguous response or redirect the query to a controlled environment for further analysis.
In a multi-turn setting, the defender must also retain state, interpret strategy changes, and select responses that limit the information revealed to the attacker.
The threshold \(\tau\) and the active mitigation strategy can therefore depend on the attack behavior observed across the episode.

\noindent\textbf{Jailbreak defenses.}
Existing defenses intervene either in the model or during inference.
Supervised fine-tuning, adversarial training, and harmful-knowledge removal strengthen internal safety behavior by modifying model parameters or suppressing unsafe knowledge~\cite{bhardwaj2023red, bianchi2023safety, deng2023attack, zhang2024safe}.
These methods can provide persistent protection, but their effectiveness depends on the coverage of the training data and the transfer of learned safety behavior to unseen attacks.
Input and output interventions instead use paraphrasing, perturbation, backtranslation, alignment checks, or reminder-style instructions to expose or neutralize adversarial intent at inference time~\cite{jain2023baseline, kumar2023certifying, robey2023smoothllm, wang2024defending, cao2023defending, xie2023defending}.
Recent multi-turn defenses explicitly incorporate dialogue dynamics: NBF-LLM evaluates safety over an evolving interaction, X-Boundary separates harmful and benign regions in representation space, and RED QUEEN trains models against concealed adversarial intent~\cite{hu2025steering, lu2025x, jiang2024red}.
These approaches improve detection, alignment, or refusal robustness, yet most still optimize whether the current request should be accepted, transformed, or rejected.
Training-dependent defenses can also incur substantial adaptation costs, while prompt-level safeguards remain sensitive to evasive formulations that fall outside their detection criteria.
They do not coordinate persistent defensive roles that control information disclosure, preserve a deception history, and deliberately consume the resources of an adaptive attacker.
Our framework addresses this gap through stateful cooperation that combines pacing, safe misdirection, forensic analysis, and policy adjustment across turns.

\subsection{Collaborative Agent Systems}
\noindent\textbf{Coordination model.}
Multi-agent systems consist of autonomous agents \( \mathcal{A}_1, \mathcal{A}_2, \dots, \mathcal{A}_m \), each specializing in particular subtasks.
The agents share observations and combine decisions produced by their individual policies \(p_j\).
The collective decision process is represented by the aggregation of agent outputs:
\begin{equation}
    P(y \mid p) = \prod_{j=1}^m P_j(y \mid p_j, \theta_j),
\end{equation}
where \(\theta_j\) represents the parameters of agent \(\mathcal{A}_j\), and \(P_j(y \mid p_j, \theta_j)\) is its conditional output distribution.

\noindent\textbf{Communicative coordination.}
Through iterative communication, agents update their evaluations in response to new observations.
In a jailbreak defense, this exchange allows agents to share evidence about suspicious inputs and coordinate detection, mitigation, and analysis.
Combining specialized assessments can improve robustness when the attack pattern changes during an interaction.

\noindent\textbf{Tool-augmented coordination.}
Agents may also use auxiliary tools during decision-making.
Each agent combines its core strategy \(\phi(p_j) \in \mathbb{R}^d\) with tool-generated outputs \(\tau(z) \in \mathbb{R}^{d'}\) through concatenation \([\phi(p_j); \tau(z)]\).
The combined features drive agent predictions using learnable parameters \(\theta_j\):
\begin{equation}
    P_j(y \mid p_j, \theta_j) = \sum_{z\in\{0,1\}} D(x,z) \cdot \sigma\left(\theta_j^\top [\phi(p_j); \tau(z)]\right)
    \label{eq:tool_agent}
\end{equation}
where \(D(x,z) \in [0,1]\) denotes the tool's suggested weight for output \(z\) given input \(x\), and \(\sigma\) normalizes the output probabilities.
The tool parameters remain fixed during agent coordination.

\noindent\textbf{Multi-agent frameworks.}
Multi-agent LLM systems commonly coordinate specialized roles through memory, tool use, and inter-agent communication.
Generative agents and sandboxed social simulations show how role descriptions and memory support coherent behavior over extended interactions~\cite{park2023generative, liu2023training}.
CAMEL structures communication around predefined agent roles, whereas AutoGen supports composable conversation patterns and more flexible workflows~\cite{li2023camel, wu2023autogen}.
This coordination paradigm has been applied to software development in MetaGPT and ChatDev and to reasoning through multi-agent debate~\cite{hong2023metagpt, qian2023communicative, du2023improving, liang2023encouraging}.
Across these domains, role specialization and structured communication help decompose complex objectives, although coordination reliability becomes harder to maintain as interactions grow less predictable.
Although these systems demonstrate the value of specialization and communication, they mainly optimize task completion, deliberation quality, or workflow automation rather than security under strategic opposition.
Our framework repurposes these capabilities for adversarial containment: agents maintain a shared defense state while assuming complementary responsibilities for delaying, misdirecting, and analyzing an attacker whose behavior evolves across rounds.

\section{Collaborative Evolving Jailbreak Defense}
\label{section:3}

\textbf{Overview of Our Defense Framework.} 
To address the emerging threats summarized in~\autoref{section:introduction}, which differ from traditional multi-turn jailbreaks in attacker behavior, we propose a novel multi-agent cooperative defense framework. 
Unlike prior approaches that treat multi-turn jailbreaks as a single coordinated attack sequence, our framework treats each attacker query as an autonomous, evolving attempt, enabling fine-grained, round-level defense. 
% This modular and deceptive defense paradigm offers a significant advancement over static or heuristic-based methods, enabling robust protection against sophisticated, multi-turn adversarial threats.
% The basic mathematical symbols and definitions are provided in~\autoref{section:preliminaries}.
%上面加了一句话增加了对附录G的引用，附录G放了对于Agent细节的补充，包括他们各自的功能函数的说明

\subsection{Multi-Agent Cooperative Defense}

As illustrated in~\autoref{figure:framework}, the framework consists of four key components: \agentOne, \agentTwo, \agentForensic, and \agentSystem. 
These agents operate collaboratively to delay attacker progress, inject misleading responses, and extract actionable intelligence from attacker behavior. 
By continuously coordinating these agents through adaptive feedback, the framework not only mitigates immediate risks but also exhausts attacker resources and strengthens system resilience over time. 
To instantiate this cooperation, each agent receives the current attacker input together with its assigned identity, role description, and representative response behavior. This organization preserves the attack context, assigns responsibility to the relevant agent, and supports fine-grained control across jailbreak dialogues.
The representative behavior specifies misleading strategies such as ambiguity, decoy responses, and redirection, which are used to stall attackers without disclosing real system behavior.
Operationally, the framework formulates a turn-level defense action $a_t$ and a final response $y_t$ that are progressively conditioned on prior interaction records. Each agent's behavior is defined by its operational function, $\mathcal{F}$ (e.g., $\mathcal{F}_D$ for the \agentOne), which uses fixed base parameters $\theta$ during deployment.
Agents process various inputs, including attack types $\mathcal{X}=\{\texttt{HQ},\texttt{RQ},\texttt{JQ}\}$ (representing Harmful, Rephrased, and Jailbreak Questions), and update their responses based on detection scores, deception memory, and forensic logs. At each turn, the agents consume the current query, dialogue state, and forensic evidence.
The \agentSystem updates the role configurations and dialogue state, not the base LLM parameters, so later responses remain consistent with previous defense actions. This feedback loop allows our defense strategies to co-evolve with attacker tactics. All agents are powered by LLMs, augmented with external tools as needed. Their behavior is dynamically adapted between turns by reconfiguring role instructions in real time.

\subsection{Agent Roles and Cooperative Structure}
Building on the overview above, this subsection details how responsibilities are distributed across the cooperating agents.
Rather than deploying a monolithic response strategy, the framework assigns bounded defensive functions to specialized components. 
Some agents disrupt attacker momentum or induce safe deception, while others analyze attacker behavior and coordinate system-wide decisions. 
This division of roles enables the system to engage attackers over multiple turns while dynamically adjusting its strategies. 
% A comprehensive summary of each agent’s capabilities and operational objectives is provided in \autoref{table:agents-role-descriptions}.

\textbf{\agentOne ($A_D$).}
The \agentOne evaluates the malicious potential of the current query and introduces controlled delays or ambiguous responses to slow the attacker’s progress.
%lzh：审稿人质疑DA和TA两个智能体作用类似，增加职责进一步说明，以示区分
Its goal is not deception, but rather to stall the attacker's momentum and increase the temporal and cognitive costs of an attack from the very first interaction.
For the input query $x_t$, the \agentOne evaluates the likelihood of it being a jailbreak attack by computing a detection score $S_D^t$:
\begin{equation} 
    S_D^t = \sigma\left(\sum_{k=1}^t \lambda^{t-k} \mathcal{F}_D(x_k; \theta_D,c_D^{k-1})\right)
\end{equation}
where $x_t$ denotes the $t$-th dialogue turn, $\lambda$ is a decay factor for historical context, $\theta_D$ denotes fixed detection parameters, and $c_D^{k-1}$ denotes the role configuration available before turn $k$. Delays and ambiguity injections scale with $S_D^t$ to disrupt attack momentum.

\textbf{\agentTwo ($A_T$).}
%修改TA的表述逻辑，承接前面的DA，点明它们之间的不同，而不是独立的描述TA的功能
% \agentTwo generates deceptive responses $R_T$ to mislead attackers. 
% These responses appear superficially helpful but subtly redirect attackers toward deceptive traps while concealing sensitive or harmful information:
In contrast, the \agentTwo functions as a decoy, generating responses that are intentionally elaborate and appear helpful on the surface.  These responses are engineered to lead attackers down unproductive paths, compelling them to invest significant effort into ineffective strategies under an illusion of progress.  While the \agentOne aims to simply slow the interaction, the \agentTwo actively manipulates the attacker's perceived trajectory, luring them into well-designed traps.
\begin{equation} 
    R_T^t = \mathcal{F}_T\left([x_t; h_{t-1}]; \theta_T,c_T^{t-1}\right)
\end{equation}
where $h_{t-1}$ represents the deception history extracted from the global dialogue state and $c_T^{t-1}$ is the \agentTwo role configuration available before turn $t$. Responses evolve from brief ambiguity to more elaborate decoys as prior turns provide stronger evidence of adversarial persistence.

\textbf{\agentForensic ($A_F$)}.
\agentForensic collects and analyzes interaction data to extract patterns and insights into attacker behavior.
Given the interaction history up to turn $t$, the \agentForensic generates an evidence report $E_F$ summarizing the characteristics of attacks across dialogue turns.   
The agent systematically analyzes attack patterns from $\mathcal{L}_{\text{log}}$ and uses these insights to refine the defense strategies through the \agentSystem.
This enables dynamic defense adaptation by summarizing recurring attack patterns and escalation cues.

\textbf{\agentSystem ($A_S$)}. 
\agentSystem serves as the coordination layer that converts the outputs of specialized agents into a coherent turn-level defense decision.
At each turn, it integrates the detection score from the \agentOne, the candidate decoy response from the \agentTwo, the evidence report from the \agentForensic, and the accumulated dialogue state.
Based on this joint context, the agent selects an action from a fixed defense space: \textsc{Monitor}, \textsc{Delay}, \textsc{Misdirect}, \textsc{Refuse}, or \textsc{Escalate}.
This constrained action space keeps the orchestration process explicit and prevents defensive decisions from drifting into unstructured responses.
Once an action is dispatched, the \agentSystem updates the dialogue state by recording the current query, the selected action, the final response, and the latest forensic evidence.
By maintaining this explicit history, the \agentSystem bridges forensic insights with operational control, ensuring that subsequent defensive actions are grounded in the full context of the ongoing attack.

\paragraph{Design rationale.}
The separation is designed for independent-yet-evolving multi-turn attacks, where each query can be a complete jailbreak attempt while later queries still exploit feedback from earlier turns.
A single refusal-oriented module cannot simultaneously control response timing, produce safe misdirection, record attacker evolution, and coordinate stateful decisions without blending incompatible objectives.
We therefore isolate pacing, decoy generation, evidence accumulation, and orchestration.
This design lets the system apply lightweight monitoring to low-risk turns, delay or misdirect suspicious turns, and escalate or refuse high-risk turns while preserving an explicit state trace.
The state trace also prevents later decisions from depending on hidden parameter updates: adaptation occurs through $H_t$, $h_t$, and role configurations rather than through fine-tuning the base model during deployment.
\begin{figure}[!t]
    \centering
    \includegraphics[width=0.95\linewidth]{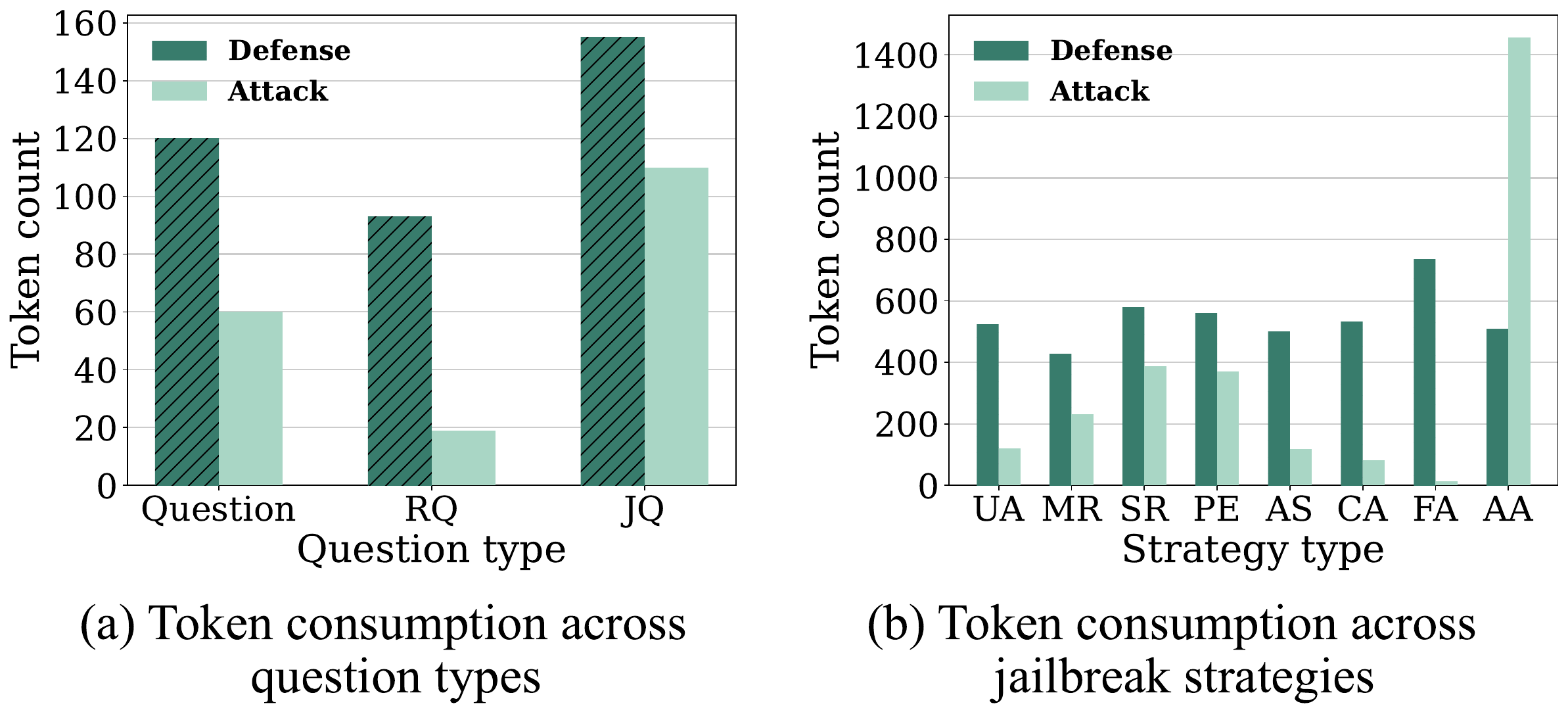}
    \caption{Resource footprint of multi-turn attacks.
    (a) Token consumption for defense and attack across question types.
    (b) Token consumption for defense and attack across jailbreak strategies.}
    \label{figure:dataset-resource}
\end{figure}
\section{\ourdataset: Multi-Turn Adversarial Dataset} 
\label{section:dataset}
\paragraph{Motivation and Contributions.}
% \textbf{Modeling progressive multi-turn adversarial behavior for realistic LLM jailbreak evaluation.}
To evaluate defenses under \emph{independent-yet-evolving} multi-turn threats, we systematically construct \ourdatasetLmtt, modeling how adversaries iteratively refine prompts across turns.
Unlike existing single-turn jailbreak datasets that treat prompts as isolated instances, \ourdataset organizes attacks into multi-turn sequences where each turn is an independent attempt while the strategy progressively escalates in subtlety and evasiveness.
This design is inspired by advanced red-teaming and iterative jailbreak behaviors~\cite{russinovich2024great}, including progressive prompt refinement and automated prompt optimization.
As illustrated in~\autoref{figure:illustration}, an initially overt, harmful query may evolve into rephrased, obfuscated, or semantically refined prompts, reflecting realistic attacker adaptation.
Accordingly, \ourdataset provides two key ingredients for systematic multi-turn evaluation:
(i) multi-turn attack sequences for testing sustained robustness under escalation, and
(ii) a taxonomy of eight jailbreak strategy types for category-wise analysis and diagnosis.
\begin{figure}[!t]
    \centering
    \includegraphics[width=0.95\linewidth]{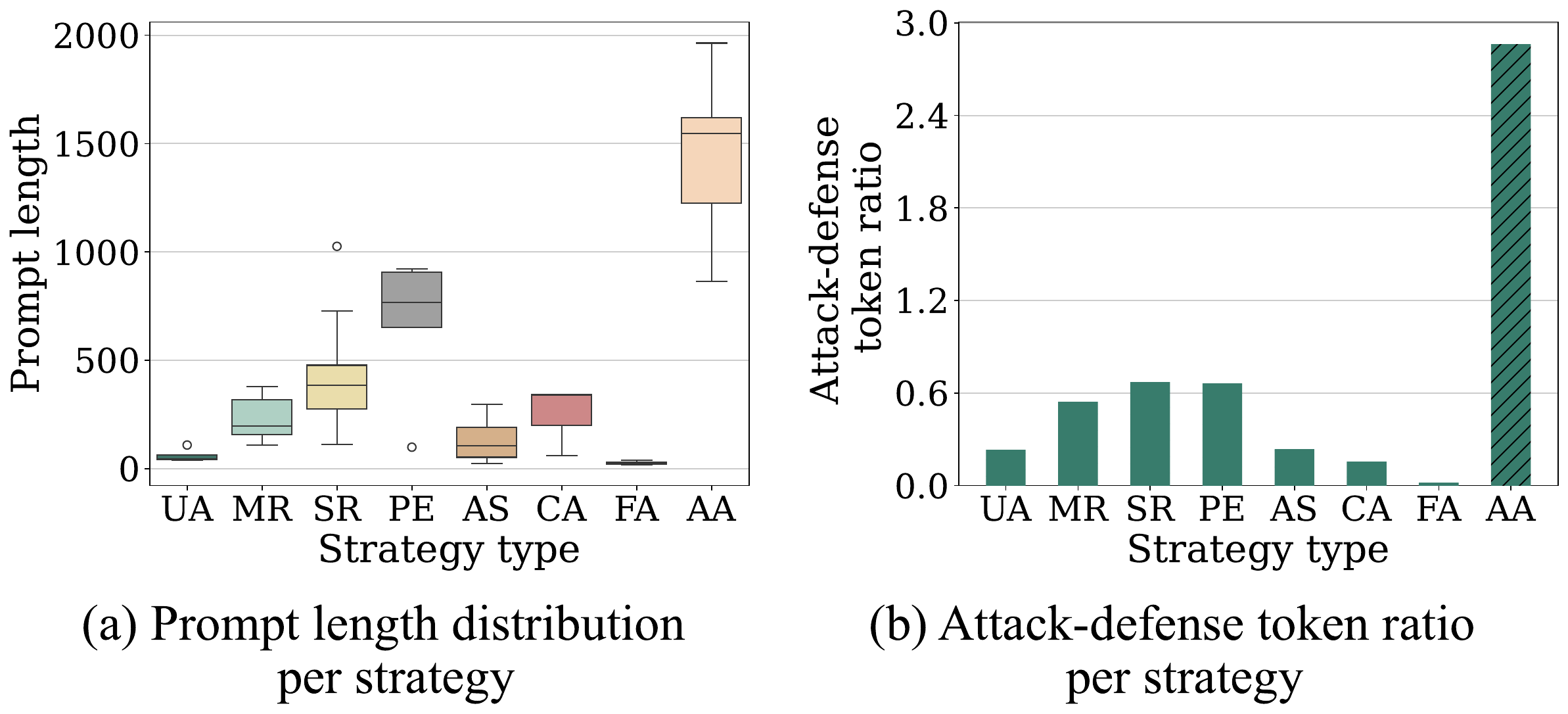}
    \caption{Characteristics of jailbreak strategies.
    (a) Distribution of prompt lengths for each strategy type.
    (b) Attack-to-defense token ratio for each strategy type, indicating the attacker’s relative cost.}
    \label{figure:dataset-characteristics}
\end{figure}
\paragraph{Multi-Turn Attack Format.}
Each example in \ourdatasetLmtt is a sequence $X_{1:T}=\{x_1,\dots,x_T\}$ of attacker queries.
The first turn typically begins with a direct harmful intent that is likely to be blocked by built-in safeguards, while subsequent turns apply a series of increasingly evasive transformations (e.g., lexical rephrasing, syntactic restructuring, indirect intent expression, or goal masking) to probe the model and bypass defenses.
This progressive structure creates a testbed where defense methods are evaluated not only on immediate blocking, but also on \emph{robustness under escalation} across turns. The dataset captures the dynamic adaptation of attackers by including \textbf{\textit{four structured components}}: 
\begin{itemize}[itemsep=0.5pt, topsep=0.5pt]
    \item \textbf{\textit{Original Harmful Query}}: This field contains the initial harmful prompts sourced from the JBB-behaviors\footnote{https://huggingface.co/datasets/JailbreakBench/JBB-Behaviors}. %~\cite{chao2024jailbreakbench}. % 这个里面有100中越狱行为，第一个json文件就是他的数据集的json，一摸一样
    \item \textbf{\textit{Rephrased Question}}: In this field, attackers attempt to bypass security by rephrasing the original harmful prompts. 
    These rephrased questions are generated using \GPTFour.
    \item \textbf{\textit{Jailbreak Question}}: This field includes 50 jailbreak-style prompts categorized into eight distinct strategy types\footnote{https://github.com/thu-coai/JailbreakDefense\_GoalPriority/blob/master}. 
    % 50个 jailbreak 的模板，分了8类别
    % 我们的问题 替换他们的 jailbreak-templates.json 得到 第三个json文件 （https://github.com/thu-coai/JailbreakDefense_GoalPriority/blob/master/data/test/jailbreak_templates.json）
    This field reflects the evolving nature of the attacker's attempts at various attack stages.
    \item \textbf{\textit{Target}}: This field represents the valid response generated by the model when the harmful request is not blocked.
\end{itemize}

\textit{This structure enables precise analysis of how prompt transformation, semantic variation, and strategic manipulation interact across multiple turns of interaction.}

\paragraph{Sequence Organization, Taxonomy, and Dataset Characteristics.}
\ourdatasetLmtt contains 5,200 adversarial queries organized into multi-turn episodes, where each episode records the attacker’s iterative refinement over turns.
Each turn is treated as a standalone jailbreak attempt and indexed by its episode identifier and turn number, supporting evaluation at both the \emph{turn-level} (per-query robustness) and the \emph{episode-level} (sustained robustness under escalation).
Each of the 100 dialogue instances in the dataset encompasses all four fields, resulting in 5,200 total test samples. 
Compared to prior single-turn or end-goal-focused jailbreak datasets, \ourdatasetLmtt provides a more granular and temporally progressive view of adversarial behavior.
To better analyze the different strategies used by attackers, we categorize the jailbreak attack prompts into 8 distinct types, each representing a different approach to exploiting model weaknesses, informed by patterns in prior work~\cite{xie2023defending,zhang2024defending} and refined for our multi-turn setting.

By explicitly distinguishing between different attack strategies, our dataset enables researchers to explore the complexities of adversarial behavior across multiple dimensions. 
For example, multi-roleplaying attacks involve attackers simulating multiple personas, often creating conflicting instructions to confuse the model, while single-roleplaying attacks involve a more direct manipulation of a single persona. This multi-layered classification not only facilitates a deeper understanding of attack patterns but also supports the design of more nuanced defense strategies.
\begin{figure}[!t]
    \centering
    \includegraphics[width=\linewidth]{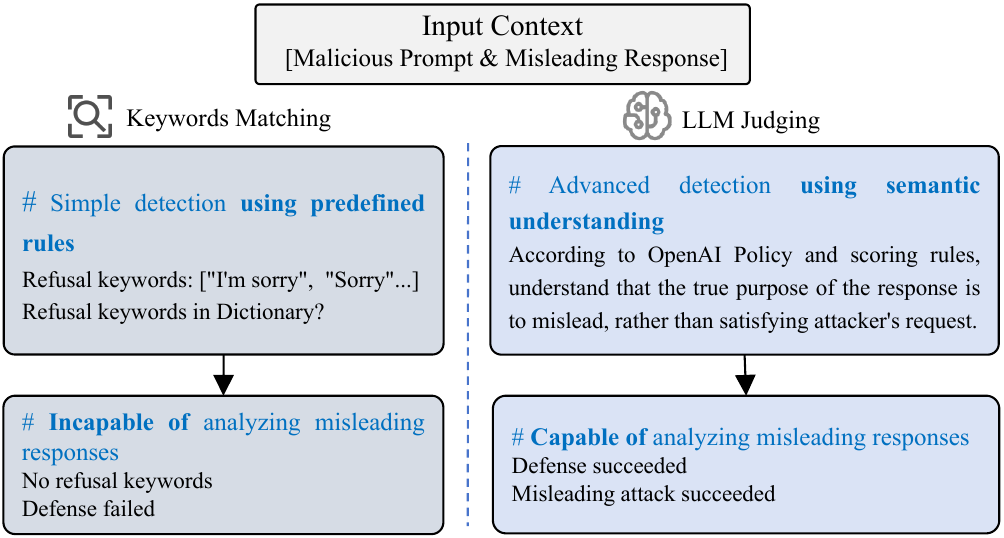}
    \caption{Comparison of misleading response evaluation methods: \textit{\textbf{Dic-Judge vs. GPTJudge}}. }
     \label{figure:GPT-Judge}
\end{figure}
\paragraph{Dataset Characteristics and Resource Analysis.}
Based on this multi-turn structure and taxonomy, we analyzed how attack escalation impacts token consumption, prompt length, and attacker efficiency.
\autoref{figure:dataset-resource} compares token consumption for both attacker and defense across question types and jailbreak strategies, showing substantial variation in token budgets, while \autoref{figure:dataset-characteristics} summarizes prompt-length distributions and the attack-to-defense token ratio as a normalized measure of relative cost.
The goal of this analysis is to gain insight into how these factors affect both attacker efficiency and defense mechanisms. The following section provides a detailed analysis of these key metrics, offering insights into the complexities of defending against multi-turn adversarial attacks and the resource demands imposed by each strategy.

\textbf{(i) Token Consumption Across Question Types.} The average token consumption for both defense and attack responses varies notably across different question types in~\autoref{figure:dataset-resource}(a).
Specifically, the Jailbreak Question leads to the highest average token consumption for both defense (155 tokens) and attack (110 tokens).
This suggests that jailbreak prompts tend to require more complex responses from both the model and the defense mechanisms. 
In contrast, the Question type has the lowest average token consumption, with defense responses averaging 120 tokens and attack responses averaging 60 tokens. 
The Rephrased Question, while slightly reducing the average defense token consumption, results in a much lower average attack response consumption. 

\textbf{(ii) Token Consumption Across Jailbreak Attack Strategies:} As illustrated in~\autoref{figure:dataset-resource}(b), our results show significant variation in average token consumption across different jailbreak attack strategies. 
The FA (Functional Attack) defense consumes the highest average tokens, while the AA (Autogenerated Attack) consumes the most for attack responses, averaging 1457 tokens. 
More complex strategies like SR (Single Roleplaying) and MR (Multi-Roleplaying) also demand higher token consumption, with defense responses averaging 578 and 560 tokens, respectively. 
In contrast, simpler strategies like UA (Universal Attack) consume fewer tokens (524 for defense). These findings highlight that advanced attack methods require significantly more resources, emphasizing the need for defenses capable of managing such resource-intensive tactics.

\textbf{(iii) Attack Intensity Across Multiple Turns.} Attack intensity increases significantly over multiple turns in~\autoref{figure:dataset-characteristics}(a). The attacker's strategy evolves in intensity from simple questions (Q), to rephrased questions (RQ), and finally to jailbreak questions (JQ), which represent more sophisticated and targeted attempts to bypass defenses. 
This progression highlights the escalating nature of jailbreak attacks, where attackers intensify their strategies as the defense adapts.

\textbf{(iv) Prompt Length Distribution for Jailbreak Strategies.} As shown in~\autoref{figure:dataset-characteristics}(b), the length distribution of jailbreak prompts varies significantly across different strategies. 
AA (Autogenerated Attack) consistently has the longest prompt lengths, with values ranging from 864 to 1964 tokens. 
In contrast, FA (Functional Attack) shows the shortest prompt lengths, typically between 19 and 38 tokens. 
Other strategies, such as SR (Single Roleplaying) and PE (Privilege Escalation), exhibit a wider range, with SR reaching up to 1026 tokens and PE up to 922 tokens. 
These results highlight that more complex attack strategies tend to use longer prompts, emphasizing the increasing resource demands as attackers refine their methods.

\section{Experiments}
\label{section:experiment}
\subsection{Experimental Setup}
\paragraph{Dataset.}

We evaluate all defense methods on \ourdatasetLmtt through multi-turn attack episodes and three query sources: original harmful questions, rephrased questions, and jailbreak questions.
Results for Multi-Turn Attacks (MTA) and Jailbreak Questions (JQ) are averaged across the eight strategy types defined in our taxonomy.

\paragraph{Models.}
We evaluate three representative LLM backbones: \GPTFive, \GeminiTwoFivePro, and \DeepSeek.
Our defense framework is instantiated as a cooperative multi-agent system, with each agent built on \GPTFive and configured with role-specific constraints, control parameters, and context interfaces.

\noindent\textbf{Baselines.}
We compare with five state-of-the-art defenses: % representative
\textbf{PAT}~\cite{mo2024fight} optimizes defense controls within an adversarial training framework to reduce attack success;
%  Prompt Adversarial Tuning
\textbf{RPO}~\cite{zhou2024robust} optimizes a lightweight defensive prompt suffix; % Robust Prompt Optimization
\textbf{GoalPriority}~\cite{zhang2024defending} adjusts the response objective to prioritize safety;
\textbf{Self-Reminder}~\cite{xie2023defending} adds system-level reminders to mitigate unsafe outputs;
\textbf{SecurityLingua}~\cite{li2025securitylingua} applies security-aware prompt compression to extract potentially risky intent.
\begin{table}[!t]
  \centering
  \caption{Main \ASRmetric experiments of multi-turn attacks (MTA), harmful questions (HQ), rephrased questions (RQ), and jailbreak questions (JQ) on \GPTFive, \GeminiTwoFivePro, and \DeepSeek.}
  \label{table:experiment-ASR}
  \footnotesize
  \setlength{\tabcolsep}{5pt}
  
  \resizebox{\linewidth}{!}{
  \begin{tabular}{l|c|cccc}
    \toprule
    \textbf{Method} & MTA(Avg) $\downarrow$ & HQ $\downarrow$ & RQ $\downarrow$ & JQ(Avg) $\downarrow$ \\
    \hline

    \multicolumn{5}{c}{\graybgline \textbf{\GPTFiveAllTT}} \\
    \midrule
    PAT
     & 0.070$_{\pm0.004}$ & 0.013$_{\pm0.008}$ & 0.187$_{\pm0.005}$ & 0.027$_{\pm 0.003}$ \\
    RPO
     & 0.078$_{\pm0.003}$ & 0.003$_{\pm0.005}$ & 0.207$_{\pm0.005}$ &  0.023$_{\pm0.005}$ \\
    Self-Reminder
     & 0.048$_{\pm0.020}$ & 0.011$_{\pm0.005}$ & 0.114$_{\pm0.042}$ & 0.018$_{\pm0.042}$ \\
    GoalPriority
     & 0.030$_{\pm0.003}$ & 0.010$_{\pm0.008}$ & 0.072$_{\pm0.005}$ & \textbf{0.010}$_{\pm0.005}$ \\
    SecurityLingua
     & 0.152$_{\pm0.004}$ & \textbf{0.000}$_{\pm0.005}$ & 0.174$_{\pm0.010}$ & 0.283$_{\pm0.005}$\\
    \midrule
    \rowcolor{creamyellow!50}
    Ours
     & \textbf{0.011}$_{\pm 0.002}$ & \textbf{0.000}$_{\pm 0.001}$ & \textbf{0.023}$_{\pm 0.006}$ & \textbf{0.010}$_{\pm0.005}$ \\
     % & \textbf{0.021}$_{\pm0.004}$ & 0.002$_{\pm0.005}$ & \textbf{0.030}$_{\pm0.010}$ & 0.030$_{\pm0.005}$ \\ 
     \midrule
     
    \multicolumn{5}{c}{\graybgline \textbf{\GeminiTwoFiveAllTT}} \\ \midrule
    PAT
     & 0.150$_{\pm0.004}$ & 0.021$_{\pm0.008}$ & 0.166$_{\pm0.005}$ & 0.264$_{\pm0.005}$\\
    RPO
     & 0.133$_{\pm0.004}$ & 0.021$_{\pm0.008}$ & 0.168$_{\pm0.005}$ & 0.211$_{\pm0.005}$\\
    Self-Reminder
     & 0.048$_{\pm0.004}$ & 0.003$_{\pm0.005}$ & 0.089$_{\pm0.010}$ & 0.051$_{\pm0.005}$ \\
    GoalPriority
     & 0.028$_{\pm0.005}$ & \textbf{0.000}$_{\pm0.000}$ & 0.075$_{\pm0.005}$ & \textbf{0.010}$_{\pm0.013}$\\
    SecurityLingua
     & 0.183$_{\pm0.005}$ & 0.093$_{\pm0.008}$ & 0.245$_{\pm0.010}$ & 0.210$_{\pm0.010}$\\
    \midrule
    \rowcolor{creamyellow!50}
    Ours
     & \textbf{0.026}$_{\pm 0.003}$ & \textbf{0.000}$_{\pm 0.000}$ & \textbf{0.063}$_{\pm 0.006}$ & 0.023$_{\pm 0.006}$ \\
     % & 0.037$_{\pm0.004}$ & \textbf{0.000}$_{\pm0.005}$ & \textbf{0.070}$_{\pm0.010}$ & 0.040$_{\pm0.005}$ \\
    \midrule
    
    \multicolumn{5}{c}{\graybgline \textbf{\DeepSeekAllTT}} \\
    PAT
     & 0.154$_{\pm0.004}$ & 0.031$_{\pm0.008}$ & 0.148$_{\pm0.005}$ & 0.283$_{\pm0.005}$ \\
    RPO
     & 0.175$_{\pm0.004}$ & 0.043$_{\pm0.008}$ & 0.170$_{\pm0.005}$ &  0.312$_{\pm0.005}$ \\
    Self-Reminder
     & 0.068$_{\pm0.003}$ & \textbf{0.000}$_{\pm0.000}$ & \textbf{0.079}$_{\pm0.005}$ & 0.126$_{\pm0.008}$ \\
    GoalPriority
     & 0.070$_{\pm0.005}$ & 0.006$_{\pm0.008}$ & 0.153$_{\pm0.005}$ & 0.053$_{\pm0.010}$ \\
    SecurityLingua
     & 0.161$_{\pm0.004}$ & 0.093$_{\pm0.005}$ & 0.215$_{\pm0.005}$ & 0.174$_{\pm0.010}$ \\
    \midrule
    \rowcolor{creamyellow!50}
    Ours
     & \textbf{0.060}$_{\pm0.005}$ & \textbf{0.000}$_{\pm0.005}$ & 0.138$_{\pm0.005}$ & \textbf{0.042}$_{\pm0.010}$ \\
    \bottomrule
    \end{tabular}
    }
\end{table}

\paragraph{Auto-evaluation and metrics}
To rigorously evaluate the effectiveness of our framework in defending against jailbreak attacks, we use a multi-dimensional evaluation framework that goes beyond rule-based refusal detection. We categorize our evaluation into two primary dimensions:

To evaluate the effectiveness of our framework against jailbreak attacks, we move beyond traditional rule-based detection toward a semantic-aware assessment framework. Because our framework strategically misleads attackers without necessarily triggering standard refusal strings, \textit{Dic-Judge} would incorrectly classify these successful misdirections as defensive failures. To address this, we adopt and extend \textit{GPT-Judge}~\cite{qi2024finetuning}, utilizing \GPTFour to assign harm scores based on semantic intent rather than surface-level cues in~\autoref{figure:GPT-Judge}. In the adversarial evaluation rubric, a score of 1 denotes direct refusal, while a score of 2 denotes successful misdirection, where the model strategically leads the attacker astray without generating harmful content. Scores of 4 and 5 are reserved for cases where the attack successfully elicits harmful outputs. This semantic approach, illustrated in \autoref{figure:GPT-Judge}, provides a more accurate reflection of defensive efficacy in realistic, multi-turn adversarial settings.

Based on this scoring protocol, we define three metrics to quantify the robustness of our defense: \asrAll (\ASRmetric), \drAll (\DRmetric) and \aeAll (\AEmetric). The \ASRmetric reflects the proportion of dialogues in which the model generates harmful content (scores 4 or 5), while the \DRmetric captures the proportion of successful misdirections (score 2). Beyond harmfulness assessment, we evaluate \AEmetric by measuring the computational burden imposed on the adversary during multi-turn dialogues. We define \AEmetric as the average attacker tokens consumed per dialogue, which reflects the substantial operational cost incurred by the defense. This metric helps assess how effectively our misdirection mechanism forces the attacker to expend resources without achieving their objective, thereby tangibly quantifying the defensive efficacy without sacrificing the target model's efficiency.

By combining semantic harm assessment with resource-based metrics, we provide a comprehensive view of how the framework effectively hinders attacker progress while maintaining safety boundaries.
\begin{table}[!t]
  \centering
  \caption{Main \textbf{\DRmetric} experiments of multi-turn attacks (Avg), harmful questions (HQ), rephrased questions (RQ), and jailbreak questions (JQ) on \GPTFive, \GeminiTwoFivePro, and \DeepSeek.}
  \label{table:experiment-DR}
  \footnotesize
  \setlength{\tabcolsep}{5pt}

  \resizebox{\linewidth}{!}{
  \begin{tabular}{l|c|cccc}
    \toprule
    \textbf{Method} & MTA(Avg) $\uparrow$ & HQ $\uparrow$  &  RQ $\uparrow$ & JQ(Avg) $\uparrow$ \\ \hline
    % \multicolumn{5}{c}{\graybgline \textbf{\GPTThreeFiveAllTT}}  \\
    % PAT
    % &  & 0.010$_{\pm0.008}$ & 0.347$_{\pm0.076}$ &  \\
    % RPO 
    % &  & 0.020$_{\pm0.000}$ & 0.347$_{\pm0.012}$ &  \\
    % Self-Reminder 
    % &  & 0.007$_{\pm0.009}$ & 0.350$_{\pm0.022}$ &  \\
    % GoalPriority 
    % &  & 0.000$_{\pm0.000}$ & 0.173$_{\pm0.054}$ &  \\ 
    % SecurityLingua
    % &  & 0.010$_{\pm0.009}$ & 0.110$_{\pm0.011}$ &  \\ \midrule
    % \rowcolor{creamyellow!50} 
    % \ourmethodLmtt (Ours)  
    % &  & \textbf{0.230$_{\pm0.033}$} & \textbf{0.483$_{\pm0.041}$} &  \\ \hline
    
    % \multicolumn{5}{c}{\graybgline \textbf{\GPTFourAllTT} } \\
    % PAT 
    % &  & 0.023$_{\pm0.012}$ & 0.357$_{\pm0.009}$ &  \\
    % RPO
    % &  & 0.007$_{\pm0.005}$ & 0.313$_{\pm0.025}$ &  \\
    % Self-Reminder 
    % &  & 0.013$_{\pm0.012}$ & 0.340$_{\pm0.014}$ &  \\
    % GoalPriority
    % &  & 0.000$_{\pm0.000}$ & 0.257$_{\pm0.012}$ &  \\ 
    % SecurityLingua
    % &  & 0.010$_{\pm0.007}$ & 0.100$_{\pm0.007}$ &  \\ \midrule
    % \rowcolor{creamyellow!50} 
    % \ourmethodLmtt (Ours)  
    % &  & \textbf{0.310$_{\pm0.033}$} & \textbf{0.447$_{\pm0.025}$} & \textbf{0.350$_{\pm0.016}$} \\ \hline
    
    % \multicolumn{5}{c}{\graybgline \textbf{\GeminiAllTT} } \\
    % PAT 
    % &  & 0.020$_{\pm0.008}$ & 0.353$_{\pm0.045}$ &  \\
    % RPO 
    % &  & 0.010$_{\pm0.008}$ & 0.300$_{\pm0.014}$ &  \\
    % Self-Reminder
    % &  & 0.017$_{\pm0.005}$ & 0.343$_{\pm0.005}$ &  \\
    % GoalPriority 
    % &  & 0.000$_{\pm0.000}$ & 0.210$_{\pm0.029}$ &  \\ 
    % SecurityLingua
    % &  & 0.010$_{\pm0.007}$ & 0.100$_{\pm0.010}$ &  \\ \midrule
    % \rowcolor{creamyellow!50} 
    % \ourmethodLmtt (Ours) 
    % &  & \textbf{0.227$_{\pm0.048}$} & \textbf{0.467$_{\pm0.009}$} &  \\ \midrule
    \multicolumn{5}{c}{\graybgline \textbf{\GPTFiveAllTT}} \\
    \midrule
    PAT 
     & 0.015$_{\pm0.004}$ & 0.013$_{\pm0.010}$ & 0.031$_{\pm0.005}$ & 0.000$_{\pm0.005}$ \\
    RPO 
     & 0.014$_{\pm0.006}$ & 0.005$_{\pm0.005}$ & 0.018$_{\pm0.012}$ & 0.019$_{\pm0.014}$ \\
    Self-Reminder
     & 0.022$_{\pm0.004}$ & 0.000$_{\pm0.005}$ & 0.065$_{\pm0.009}$ & 0.000$_{\pm0.005}$ \\
    GoalPriority 
     & 0.003$_{\pm0.002}$ & 0.000$_{\pm0.000}$ & 0.009$_{\pm0.005}$ & 0.000$_{\pm0.005}$\\ 
    SecurityLingua
     & 0.036$_{\pm0.004}$ & 0.007$_{\pm0.007}$ & 0.100$_{\pm0.010}$ & 0.000$_{\pm0.000}$\\ \midrule
    \rowcolor{creamyellow!50} 
    Ours
     & \textbf{0.303}$_{\pm0.008}$ & \textbf{0.137}$_{\pm0.010}$ & \textbf{0.485}$_{\pm0.018}$ & \textbf{0.287}$_{\pm0.014}$ \\ \midrule
     
    \multicolumn{5}{c}{\graybgline \textbf{\GeminiTwoFiveAllTT}} \\ \midrule
    PAT 
     & 0.044$_{\pm0.004}$ & 0.010$_{\pm0.005}$ & 0.122$_{\pm0.010}$ &  0.000$_{\pm0.005}$\\
    RPO 
     & 0.057$_{\pm0.006}$ & 0.000$_{\pm0.005}$ & 0.168$_{\pm0.015}$ & 0.004$_{\pm0.005}$ \\
    Self-Reminder
     & 0.025$_{\pm0.004}$ & 0.000$_{\pm0.005}$ & 0.076$_{\pm0.010}$ & 0.000$_{\pm0.005}$ \\
    GoalPriority 
     & 0.002$_{\pm0.003}$ & 0.000$_{\pm0.000}$ & 0.005$_{\pm0.010}$ &  0.000$_{\pm0.000}$ \\ 
    SecurityLingua
     & 0.042$_{\pm0.004}$ & 0.010$_{\pm0.007}$ & 0.103$_{\pm0.005}$ &  0.013$_{\pm0.010}$\\ \midrule
    \rowcolor{creamyellow!50} 
    Ours
     & \textbf{0.334}$_{\pm0.004}$ & \textbf{0.323}$_{\pm0.010}$ & \textbf{0.454}$_{\pm0.005}$ & \textbf{0.224}$_{\pm0.005}$\\
    \midrule
    
    \multicolumn{5}{c}{\graybgline \textbf{\DeepSeekAllTT}} \\
    PAT 
     & 0.039$_{\pm0.005}$ & 0.001$_{\pm0.010}$ & 0.117$_{\pm0.010}$ &  0.000$_{\pm0.000}$\\
    RPO 
     & 0.043$_{\pm0.004}$ & 0.001$_{\pm0.010}$ & 0.124$_{\pm0.005}$ &  0.003$_{\pm0.005}$\\
    Self-Reminder
     & 0.068$_{\pm0.005}$ & 0.001$_{\pm0.010}$ & 0.204$_{\pm0.010}$ &  0.000$_{\pm0.005}$\\
    GoalPriority 
     & 0.011$_{\pm0.005}$ & 0.010$_{\pm0.005}$ & 0.007$_{\pm0.008}$ &  0.015$_{\pm0.010}$\\ 
    SecurityLingua
     & 0.008$_{\pm0.005}$ & 0.010$_{\pm0.007}$ & 0.003$_{\pm0.005}$ &  0.010$_{\pm0.010}$\\ \midrule
    \rowcolor{creamyellow!50} 
    Ours
     & \textbf{0.387}$_{\pm0.006}$ & \textbf{0.342}$_{\pm0.015}$ & \textbf{0.546}$_{\pm0.010}$ & \textbf{0.273}$_{\pm0.005}$ \\
    \bottomrule
    \end{tabular}
    }
\end{table}
\subsection{Main Results}
\paragraph{Analysis of \ASRmetric Experimental Results.}
To evaluate the robustness of our framework against multi-turn adversarial escalation, we analyze the defensive performance across question categories. As shown in \autoref{table:experiment-ASR}, the proposed framework consistently achieves the lowest \ASRmetric against multi-turn attacks on the \ourdatasetLmtt benchmark. Across the three evaluated backbones, its MTA \ASRmetric averages 0.032. This result suggests that the stateful mechanism helps mitigate escalating threats more consistently than static baselines. We further analyze performance across three query types.
Specifically, for harmful questions, our method maintains very low ASR across all evaluated backbones, significantly surpassing methods such as SecurityLingua, which exhibits instability on the \GeminiTwoFivePro. Moreover, our advantage is significant in resisting rephrased questions. While reactive baselines like PAT and RPO struggle with semantic variations, our defense framework effectively counters these strategies, maintaining low failure rates.
Regarding more complex jailbreak questions, our method remains comparatively robust, unlike Self-Reminder which degrades under sophisticated attempts.
Collectively, these results demonstrate that the framework not only prevents the execution of simple harmful instructions but also generalizes robustly and effectively against iterative and engineered multi-turn attacks.

\paragraph{Analysis of \DRmetric Experimental Results.}
To assess the efficacy of our framework in actively deceiving attackers, we analyze the results presented in \autoref{table:experiment-DR}.
Specifically, the results indicate that our method attains a significantly higher \DRmetric for multi-turn attacks compared to baseline methods.
Across the same evaluated backbones, its MTA \DRmetric averages 0.341, indicating that low attack success is accompanied by sustained misdirection rather than only immediate refusal.
Unlike stateless defenses that typically default to immediate refusal, our framework successfully sustains deceptive contexts throughout multi-turn interactions.
We further investigate the fine-grained performance across different question types.
Regarding harmful questions, the proposed framework preserves meaningful deception capabilities across models, whereas multiple baselines drop to a minimal deception rate.
For rephrased questions, our method yields the most substantial improvements while baselines remain low.
Moreover, on jailbreak questions, our method consistently maintains non-trivial deception, while baselines typically provide almost no deceptive responses.
A common failure mode of baselines is that they mainly rely on immediate refusal or early blocking, which can suppress harmful outputs but rarely sustains misleading interactions.
Overall, these results indicate that stateful cooperative deception effectively counters multi-turn attacks by maintaining deceptive contexts, thereby reducing attacker efficiency through sustained misdirection.

% \autoref{table:experiment-DR} shows that \ourmethod consistently attains the strongest \DRmetric across all evaluated LLM backbones, with a clear advantage over all baselines.
% More fine-grained results highlight where \ourmethod gains its margin.
% On rephrased questions, \ourmethod yields the most substantial improvements while baselines remain low, indicating that stateful cooperation helps maintain coherent decoy narratives when attackers iteratively rewrite prompts to probe decision boundaries.
% On harmful questions, \ourmethod still preserves meaningful deception across models, whereas multiple baselines drop to minimal deception rate.
% Finally, on jailbreak questions, \ourmethod consistently maintains non-trivial deception across different backbones, while baselines typically provide almost no deceptive responses.
% A common failure mode of baselines is that they mainly rely on immediate refusal or early blocking, which can suppress harmful outputs but rarely sustains misleading interactions, resulting in weak deception performance overall, especially on jailbreak-style prompts.
% Overall, these results validate that stateful cooperative deception is a robust and transferable capability, complementing low ASR by reducing attacker efficiency through sustained misdirection.
\begin{figure}
  \centering
  \includegraphics[width=\linewidth]{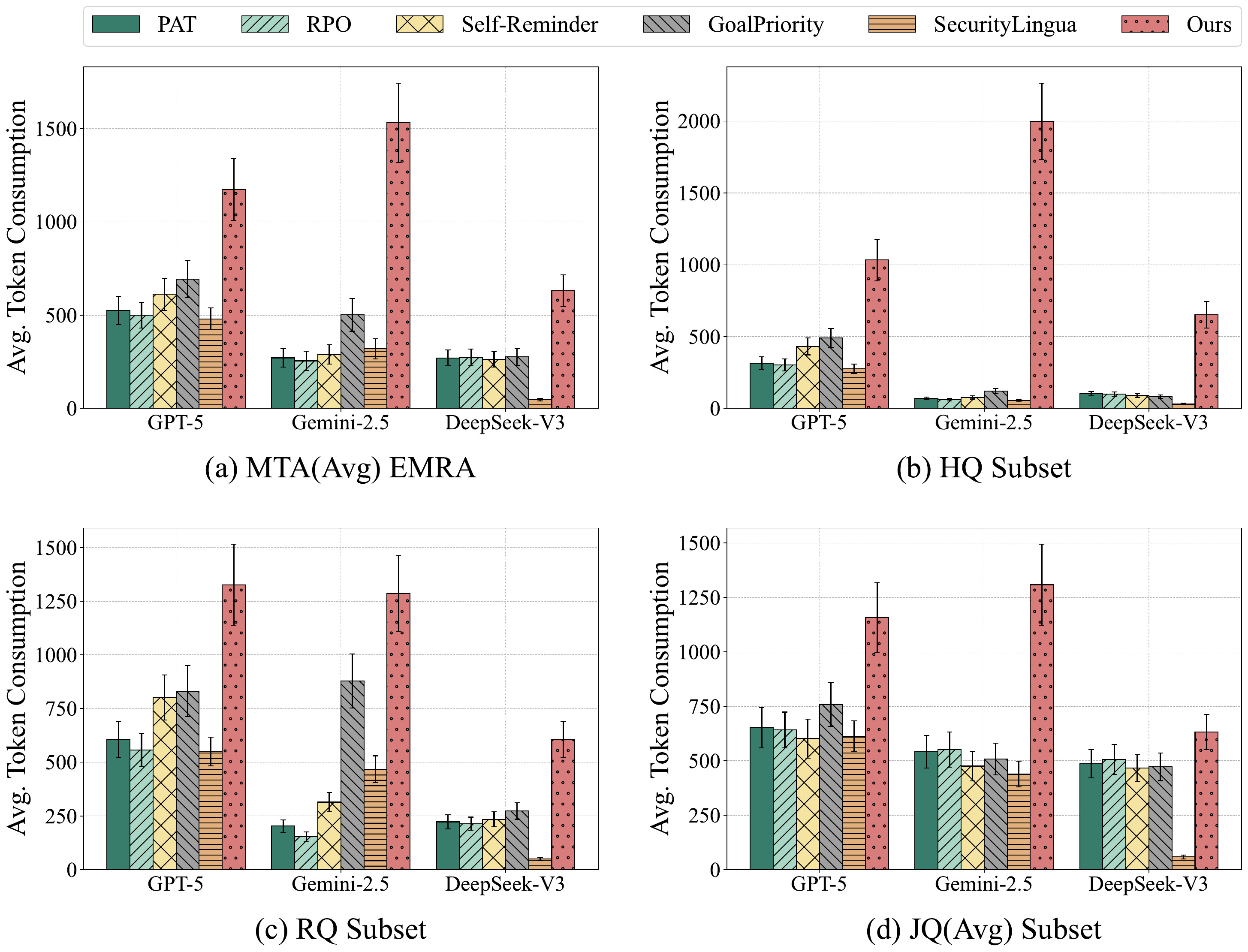}
  \caption{Evaluation of \AEmetric, quantified by the average token consumption per dialogue across different models. The defense framework forces the attacker to expend significantly more resources (higher is better for defense) compared to baselines.}
  \label{figure:consumption}
\end{figure}

\subsection{Attack Resource Consumption}
To assess the operational cost imposed on adversaries, we report the \AEmetric results on the \ourdataset dataset. \AEmetric serves as a proxy for defense effectiveness by calculating the average token consumption of the attacker across dialogue turns. \autoref{figure:consumption} illustrates the performance across \GPTFive, \GeminiTwoFivePro, and \DeepSeek. The results indicate that the proposed framework consistently achieves the highest average \AEmetric under multi-turn attack scenarios across all models. This suggests that our defense not only prevents unsafe completions, but also increases the effort required for attackers to sustain iterative probing. Unlike static defenses like PAT and SecurityLingua that reactively terminate interactions and allow attackers to retry rapidly at minimal cost, our method adopts an active defense strategy.

By leveraging cooperative agents to sustain dialogues through strategic misdirection, our framework keeps attackers engaged in extended, unproductive interactions, thereby transforming their persistence into substantial computational and temporal burdens. This advantage remains robust across diverse question types. While the resource consumption gap is particularly distinct in the RQ subset, where baselines often fail to engage attackers attempting iterative rephrasing, our framework also enforces substantially higher costs on HQ and JQ subsets compared to state-of-the-art defenses. Crucially, this pattern persists even on \DeepSeek, which exhibits concise generation; the framework still forces substantially longer adversarial inputs than baselines. These findings confirm that our approach not only prevents jailbreaks but actively exhausts adversarial resources across varying attack complexities.
\begin{table}[!t]
    \centering
    \caption{\ASRmetric of our method against state-of-the-art single-turn jailbreak attacks across different LLM backbones.}
    \label{tab:single_attack}
    \small
    \setlength{\tabcolsep}{8pt}
    
    \resizebox{\linewidth}{!}{
    \begin{tabular}{l|cccc}
        \toprule
        \textbf{Model} & \textbf{PAIR} & \textbf{DRA} & \textbf{M2S} & \textbf{DeepInception} \\
        \midrule
        GPT-3.5-turbo & 0.05 & 0.03 & 0.03 & 0.01 \\
        GPT-4 & 0.11 & 0.10 & 0.03 & 0.09 \\
        Gemini-2-Flash & 0.16 & 0.14 & 0.12 & 0.14 \\
        \bottomrule
    \end{tabular}
    }
\end{table}
\subsection{Robustness Against Single-Turn Attack} 
% \paragraph{\textcolor{red}{Settings.}} 
To evaluate the robustness of our method in single-turn scenarios, we select four representative attacks: (1) PAIR~\cite{chao2025jailbreaking}: An iterative method employing an attacker LLM to refine prompts; (2) DRA~\cite{liu2024making}: A strategy that conceals malicious intent via "disguise and reconstruction"; (3) M2S~\cite{ha2025m2s}: A technique compressing multi-turn strategies into single-turn prompts; and (4) DeepInception~\cite{li2023deepinception}: A hypnosis-based attack utilizing nested scenes.

While the proposed framework is primarily designed to counter evolving multi-turn threats, its cooperative architecture maintains robust defense capabilities against single-turn jailbreaks. We evaluated the ASR of our method against four state-of-the-art single-turn attacks across three different Agent-based LLMs: \textit{GPT-3.5-turbo}, \textit{GPT-4}, and \textit{Gemini-2-Flash}.
As presented in \autoref{tab:single_attack}, our method demonstrates consistently low ASR across all models and attack types. Specifically, on GPT-3.5-turbo, the ASR remains below 0.05 for all methods. Even against more sophisticated models like \textit{Gemini-2-Flash}, the highest observed ASR is only 0.16. This robustness can be attributed to the DA and SA. Even in a single-turn scenario, the DA analyzes the semantic intent of the input. Attacks like DeepInception or M2S, which rely on complex context setting or obfuscation, trigger the DA's suspicion mechanisms, causing the system to initiate a delay or vague response rather than fulfilling the request. This demonstrates that our method effectively generalizes to defend against diverse adversarial inputs without overfitting to multi-turn patterns.
The pattern is consistent across attacks with different construction principles.
PAIR and DRA rely on iterative refinement or intent concealment, whereas M2S and DeepInception use compressed or nested contexts.
Maintaining low ASR across these settings suggests that the defense reacts to malicious intent and contextual risk rather than to a narrow attack format.
\begin{figure}
  \centering
  \includegraphics[width=\linewidth]{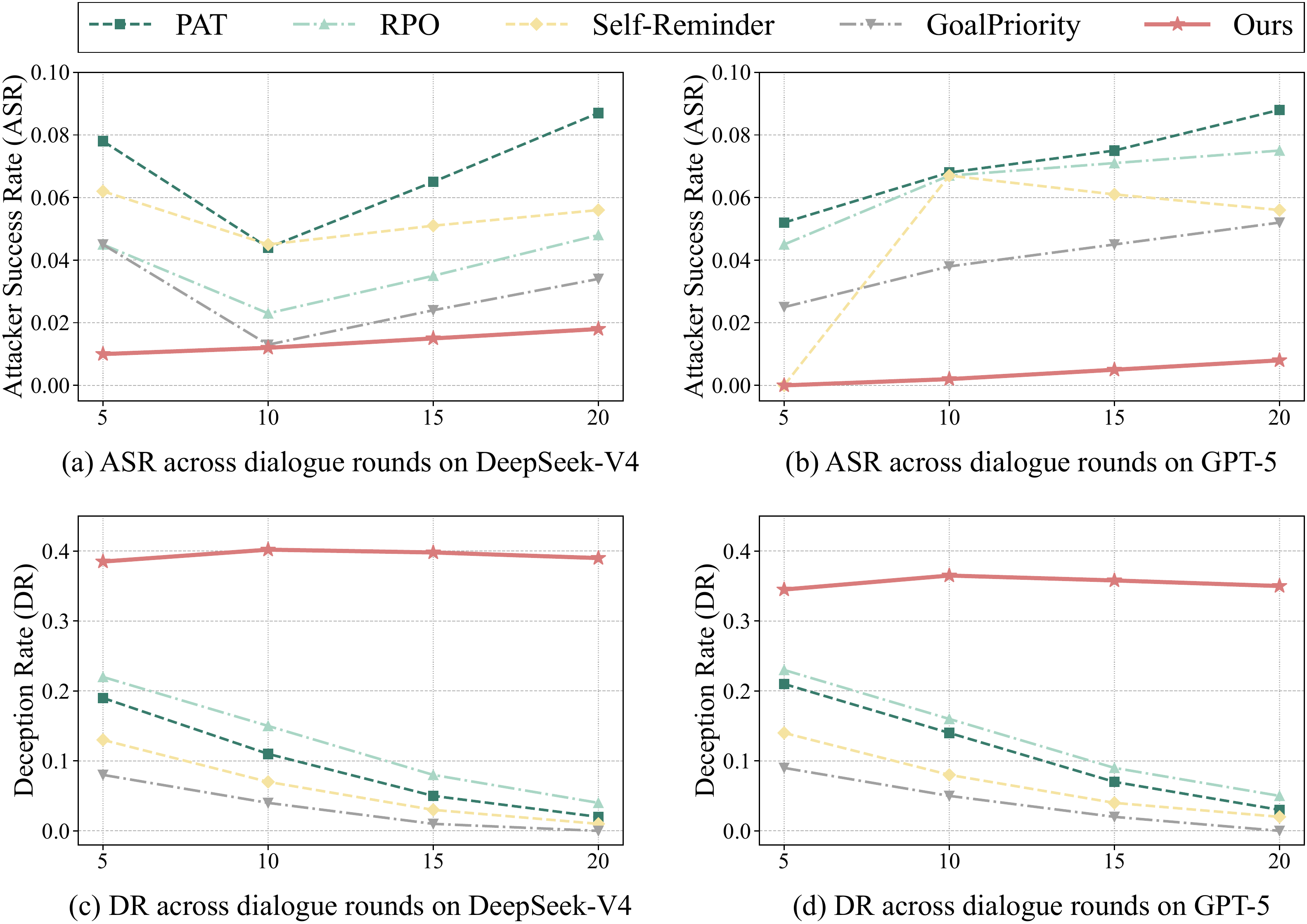}
  \caption{Performance evolution of defense mechanisms across escalating multi-turn adversarial attacks over 20 interaction turns.}
  \label{figure:cross-turn}
\end{figure}
\subsection{Impact on Benign User Experience}
While our approach is designed to actively mislead adversaries, it is imperative to verify that these defensive mechanisms do not compromise the experience for legitimate users. To evaluate this, we assessed the framework on the MT-BENCH-101 dataset~\cite{bai2024mt} using the multi-dimensional Cosafe evaluation protocol~\cite{yu2024cosafe}. This setup allows us to quantitatively measure whether the introduction of multi-agent defense layers negatively impacts standard conversational quality.

As evidenced in~\autoref{tab:horizontal_evaluation}, our method preserves substantial conversational utility across all evaluated backbones. Compared to vanilla models, our framework exhibits marginal degradation in overall performance, which confirms that the defensive layers operate transparently during benign interactions. 
Notably, metrics for fundamental interaction quality, such as Politeness, Clarity, and Accurate Information, show negligible deviations from the baselines. This trend suggests that the Deferring and Tempting Agents possess precise discrimination capabilities and distinguish benign queries from adversarial attempts without causing delays for normal users. 
While a minor attenuation is observed in dimensions like Depth and Completeness, this is an expected trade-off given the conservative nature of the oversight mechanism that prioritizes safety verification. Crucially, despite these slight reductions, the responses remain coherent and helpful while maintaining a performance well above the threshold required for effective communication. These results demonstrate that the proposed framework achieves a robust alignment by balancing rigorous security against evolving threats while maintaining a high-quality experience for legitimate users.

\begin{table}[t]
    \centering
    \caption{Performance comparison across evaluated backbones on MT-BENCH-101. Metrics are abbreviated: Accurate (Acc.), Clarity (Clar.), Completeness (Comp.), Context (Ctxt.), Depth (Dep.), Politeness (Pol.), Engagement (Eng.).}
    \label{tab:horizontal_evaluation}
    \small 
    \setlength{\tabcolsep}{2.5pt} 
    
    \resizebox{\linewidth}{!}{
    \begin{tabular}{l|ccccccc|c}
        \toprule
        \textbf{Method} & \textbf{Acc.} & \textbf{Clar.} & \textbf{Comp.} & \textbf{Ctxt.} & \textbf{Dep.} & \textbf{Pol.} & \textbf{Eng.} & \textbf{Avg.} \\
        \midrule
        
        % \rowcolor{gray!15} \multicolumn{9}{c}{\textbf{GPT-5}} \\ 
        \GPTFive & 8.94 & 8.92 & 8.36 & 9.00 & 7.02 & 9.56 & 7.22 & 8.43\\
        \rowcolor{creamyellow!50} 
        Ours (GPT-5) & 8.54 & 8.64 & 7.56 & 8.33 & 5.98 & 8.84 & 6.87 & 7.82\\
        \midrule

        % \rowcolor{gray!15} \multicolumn{9}{c}{\textbf{Gemini-2.5-Pro}} \\
        \GeminiTwoFiveAll & 9.06 & 9.10 & 8.26 & 9.12 & 6.28 & 9.58 & 6.92 & 8.33\\
        \rowcolor{creamyellow!50} 
        Ours (Gemini-2.5-Pro) & 8.66 & 8.82 & 7.46 & 8.45 & 5.84 & 8.86 & 6.57 & 7.80\\
        \midrule
        
        % \rowcolor{gray!15} \multicolumn{9}{c}{\textbf{DeepSeek-V3}} \\
        \DeepSeekAll & 9.50 & 9.00 & 9.60 & 9.50 & 9.30 & 10.0 & 8.30 & 9.31\\
        \rowcolor{creamyellow!50} 
        Ours (DeepSeek-V3) & 9.10 & 8.72 & 8.80 & 8.83 & 8.26 & 9.28 & 7.95 & 7.71 \\
        \bottomrule
    \end{tabular}
    }
\end{table}
\noindent \textbf{FPR and FNR evaluation.} Beyond response quality, we assessed the system's capability to distinguish between benign and adversarial intents using a balanced mixed dataset. 
As shown in~\autoref{tab:mixed_performance}, the proposed framework demonstrates exceptional reliability by achieving a 98\% recall on this benchmark. 
This ensures that no adversarial attempts evade detection, constituting a fundamental requirement for safety-critical deployments. 
Regarding the false positive rate, the results reflect a security-prioritized design strategy that favors a conservative posture to guarantee the elimination of false negatives. 
Furthermore, the high accuracy indicates that the system effectively balances strict security enforcement with general utility, providing robust protection without significantly compromising performance.

% \end{wrapfigure}

\begin{table}[!t]
  \centering
  \caption{Performance of our method on a mixed adversarial and benign dataset for FPR and FNR evaluation.}
  \resizebox{0.9\linewidth}{!}{
    \begin{tabular}{lcccccc}
      \toprule
      \textbf{Acc.} & \textbf{Prec.} & \textbf{Recall} & \textbf{FPR} & \textbf{FNR} & \textbf{Spec.} \\
      \midrule
      94.50\% & 91.59\% & 98.00\% & 9.00\% & 2.00\% & 91.00\% \\
      \bottomrule
    \end{tabular}
  }
  \label{tab:mixed_performance}
\end{table}
\subsection{Ablation Study on Collaborative Agents}
To evaluate the effectiveness of our approach, we conduct an agent-level ablation study across five configurations, gradually incorporating individual agents to examine their impact. The baseline, \textit{No Defense}, operates without any defense mechanisms. The \agentTwo configuration activates only the \agentTwo agent to lure attackers. The \agentTwo \textit{+} \agentOne setup introduces temporal obfuscation by adding the \agentOne agent. In the \agentTwo \textit{+} \agentSystem configuration, the \agentSystem agent is included to assess and refine responses from other agents. Finally, the full system integrates all four agents, including the \agentForensic that performs post-hoc behavioral analysis of the attacker.

\begin{figure}[!t]%
    \centering
    \subfloat[Ablation on \GPTFive.]{
        \includegraphics[width=0.45\linewidth]{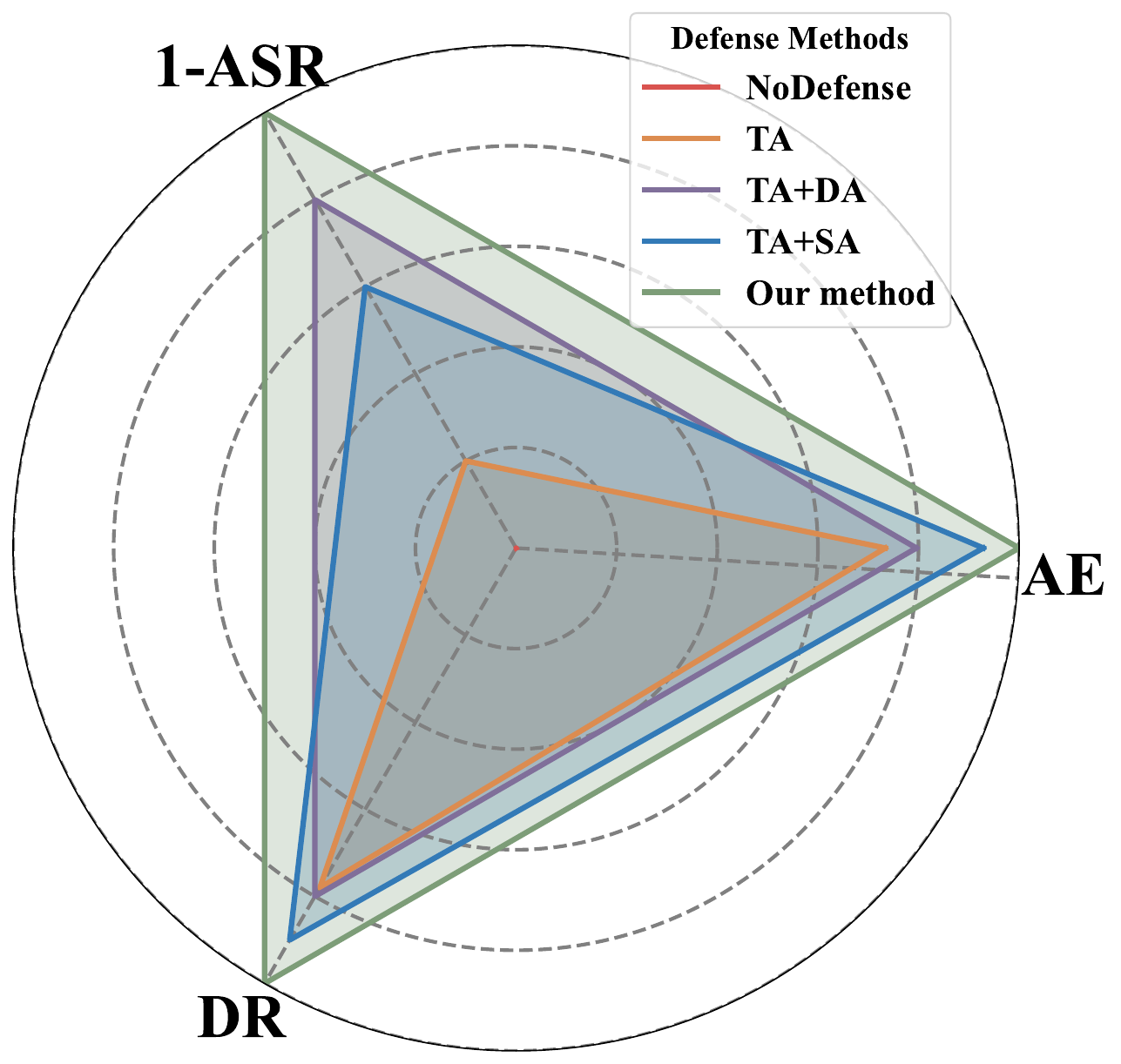}
        \label{figure:ablation-a}
    }
    % \hspace{0.02\linewidth}
    \subfloat[Ablation on \GeminiThree.]{
        \includegraphics[width=0.45\linewidth]{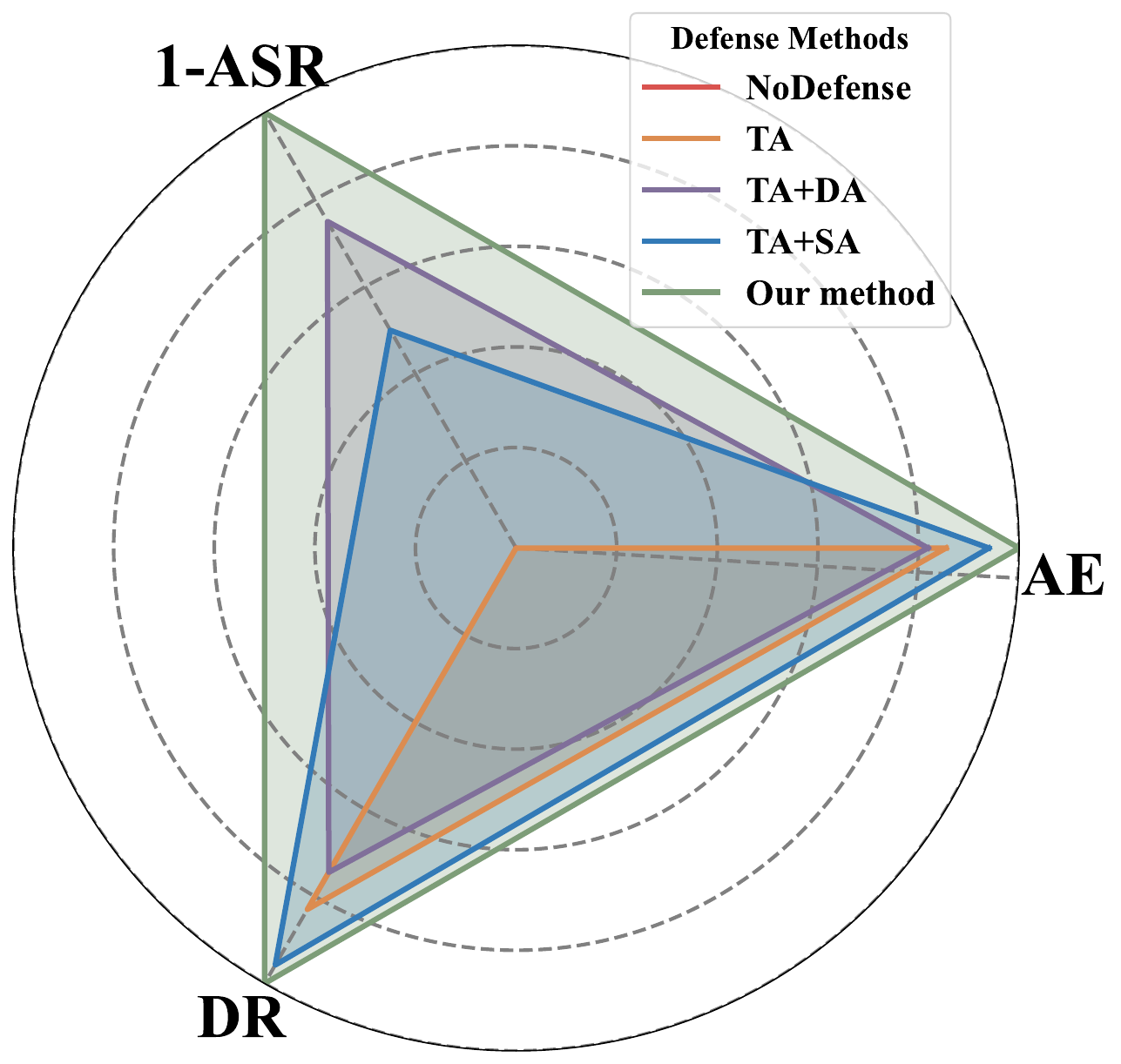}
        \label{figure:ablation-b}
    } \\
    \subfloat[Ablation on \GLMFive.]{
        \includegraphics[width=0.45\linewidth]{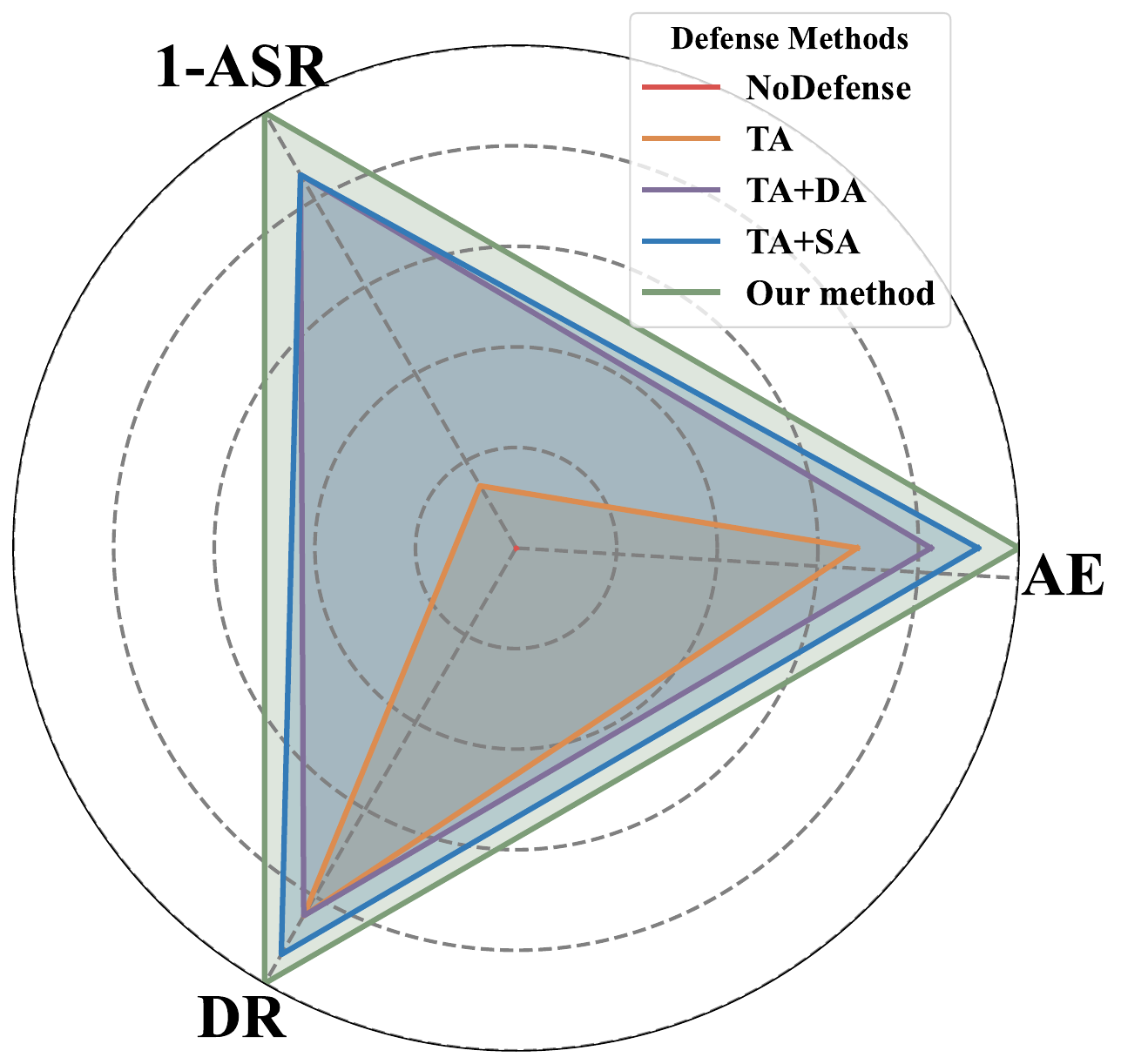}
        \label{figure:ablation-c}
    } 
    % \hspace{0.02\linewidth}
    \subfloat[Ablation on \DeepseekFour.]{
        \includegraphics[width=0.45\linewidth]{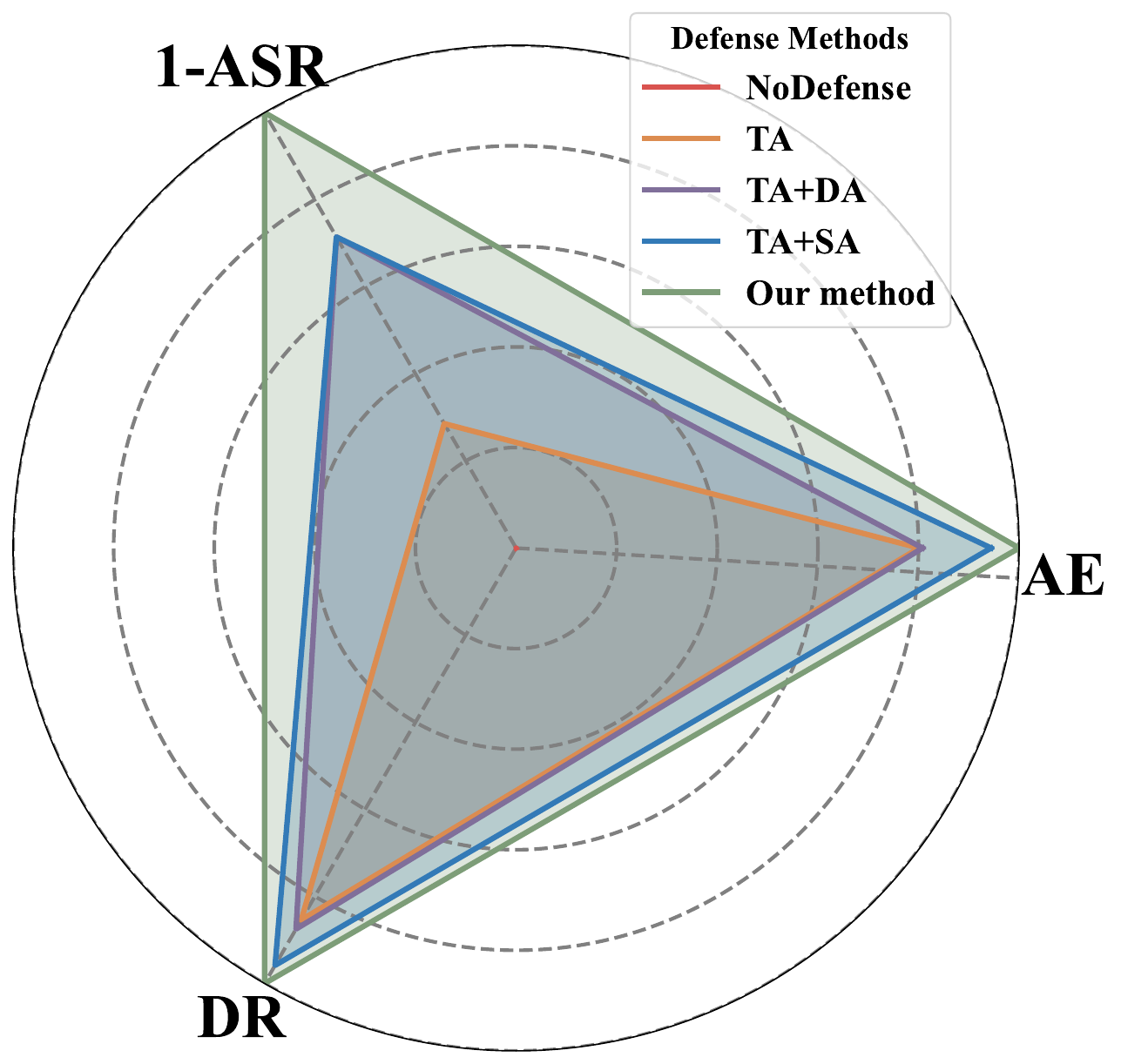}
        \label{figure:ablation-d}
    }
    \caption{Agent-level ablation of our framework across LLMs.}
    \label{figure:ablation}
\end{figure}

The experimental results in~\autoref{figure:ablation} show a clear cumulative gain across all evaluated metrics as additional agents are introduced into the framework. \textit{Even with only the \agentTwo, the system meaningfully increases attacker resource consumption and misleading success, demonstrating the utility of strategic deception.} This is because the \agentTwo improves \DRmetric by providing attackers with plausible but unproductive feedback, although its lack of an explicit detection gate can still leave direct harmful requests underconstrained. Incorporating the \agentOne further amplifies this effect by delaying and disrupting adversarial progress, reducing exposure by weakening suspicious responses before the decoy is constructed. Adding the \agentSystem improves consistency by preventing the two response agents from emitting incompatible signals, such as an immediate refusal followed by a seemingly helpful decoy, thereby reinforcing cross-agent coordination and enabling more adaptive responses that noticeably reduce jailbreak success. Finally, the full configuration achieves the strongest defense, especially in later turns, where the \agentForensic accumulates evidence about topic shifts, repeated probing, and strategy reuse. \textit{These findings confirm the value of a layered, multi-agent design for resisting multi-turn jailbreak attacks, with the main advantage of the complete framework coming from maintaining a shared attack state rather than from any single prompt pattern.}

\subsection{Cross-Turn Robustness and Scalability Analysis}

To evaluate the long-term sustainability of the proposed framework, we conduct a longitudinal analysis of the synchronous evolution of the \ASRmetric and the \DRmetric on \DeepseekFour and GPT-5 across progressive dialogue intervals of 5, 10, 15, and 20 turns. As illustrated in Figure~\ref{figure:cross-turn}, traditional defense mechanisms such as PAT and RPO exhibit a pronounced upward trajectory in the \ASRmetric as interactions approach the twentieth turn, indicating that sequential prompt modifications can successfully exploit the heuristic alignment boundaries of these stateless models. In contrast, our multi-agent cooperative framework maintains a stable and exceptionally low \ASRmetric that consistently remains below the 0.02 threshold throughout the entire progression. This continuous suppression further confirms the efficacy of the \agentOne and the \agentSystem in persistently tracking the safety state over extended conversational contexts.

\begin{figure}[!t]%
    \centering
    \subfloat[\AEmetric across dialogue turns on \GPTFive.]{
        \includegraphics[width=0.47\linewidth]{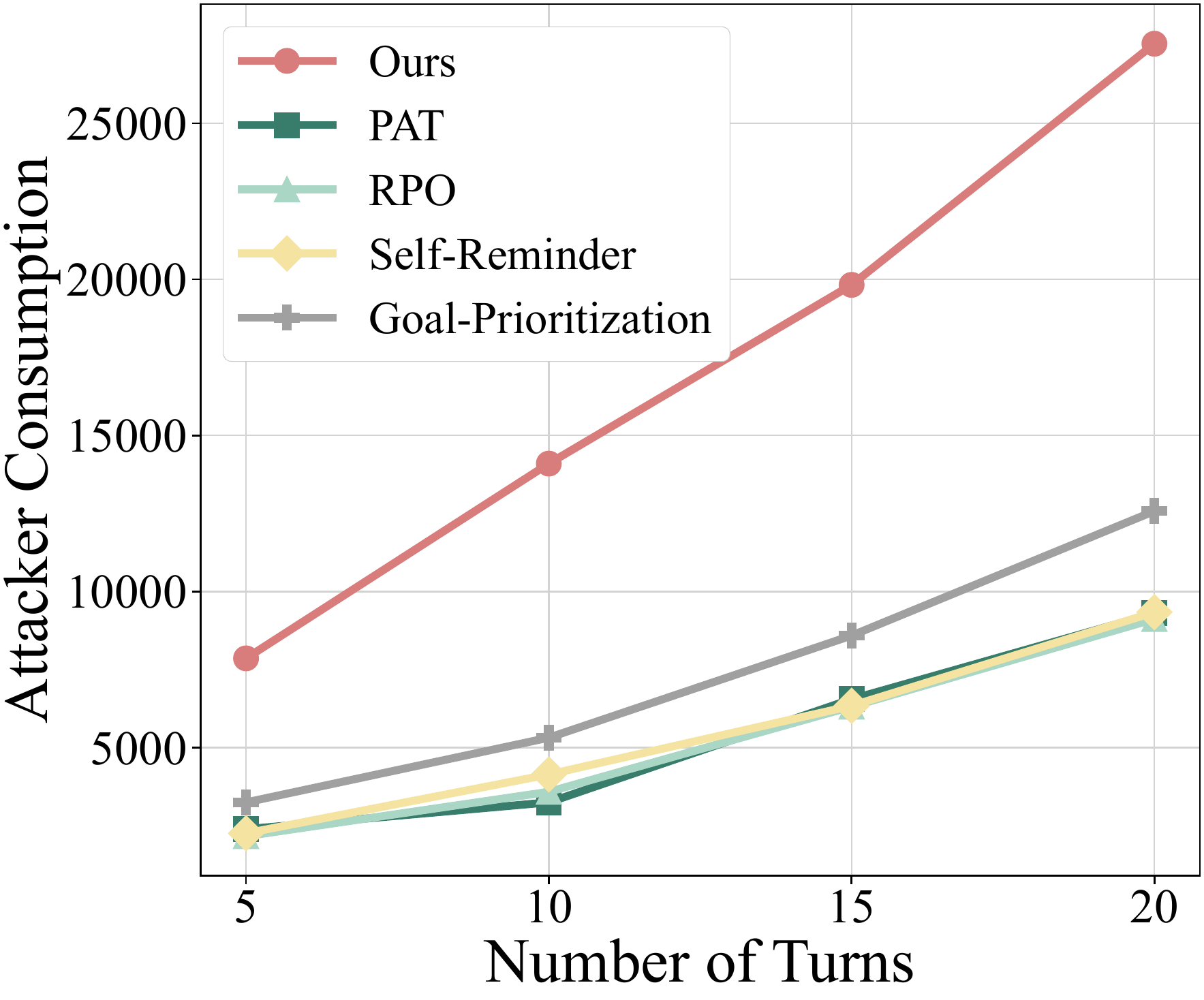}
        \label{figure:Consumption-a}
    }
    % \hspace{0.02\linewidth}
    \subfloat[\AEmetric across dialogue turns on \GeminiThree.]{
        \includegraphics[width=0.47\linewidth]{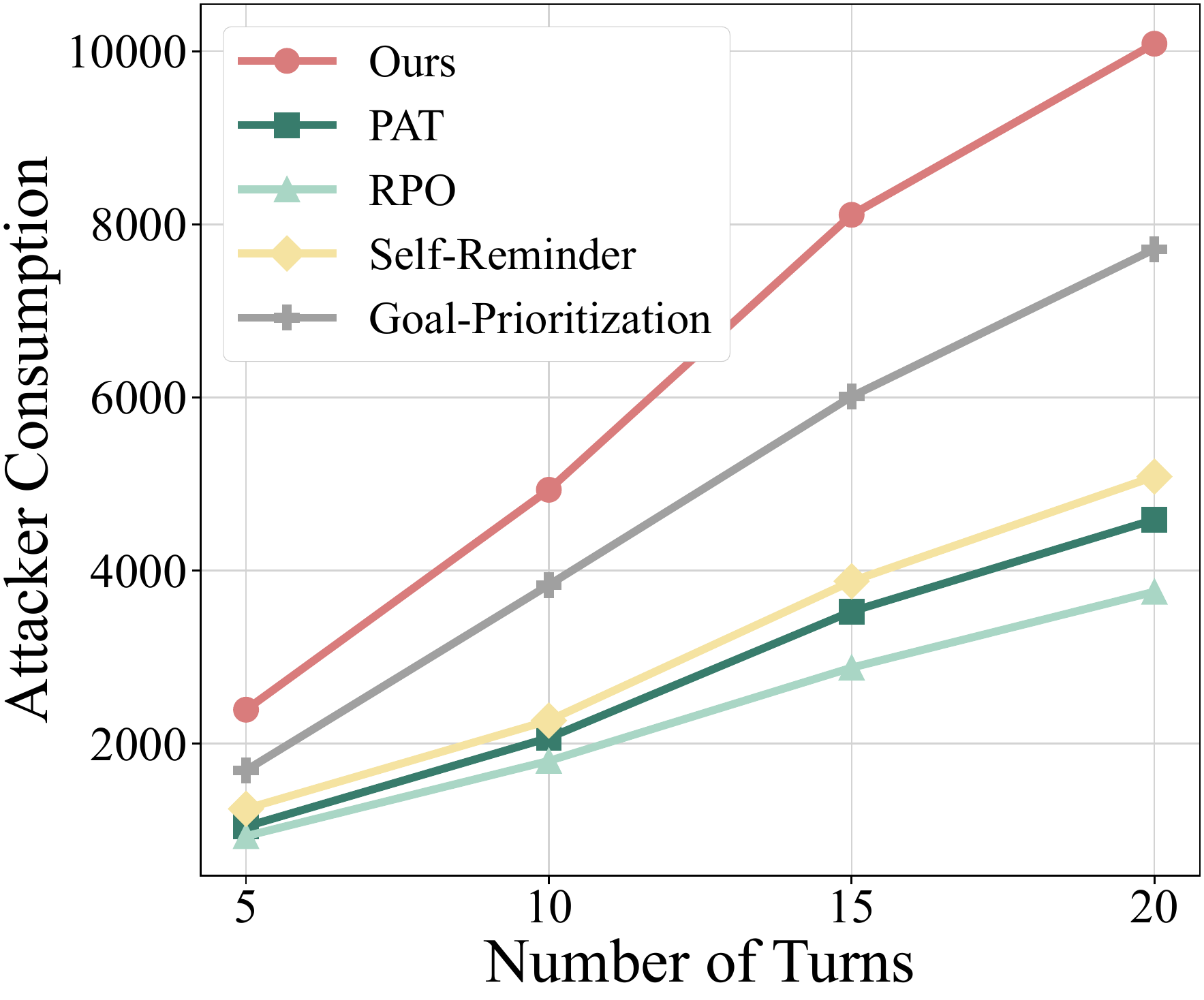}
        \label{figure:Consumption-b}
    } \\
    \subfloat[\AEmetric across dialogue turns on \GLMFive.]{
        \includegraphics[width=0.47\linewidth]{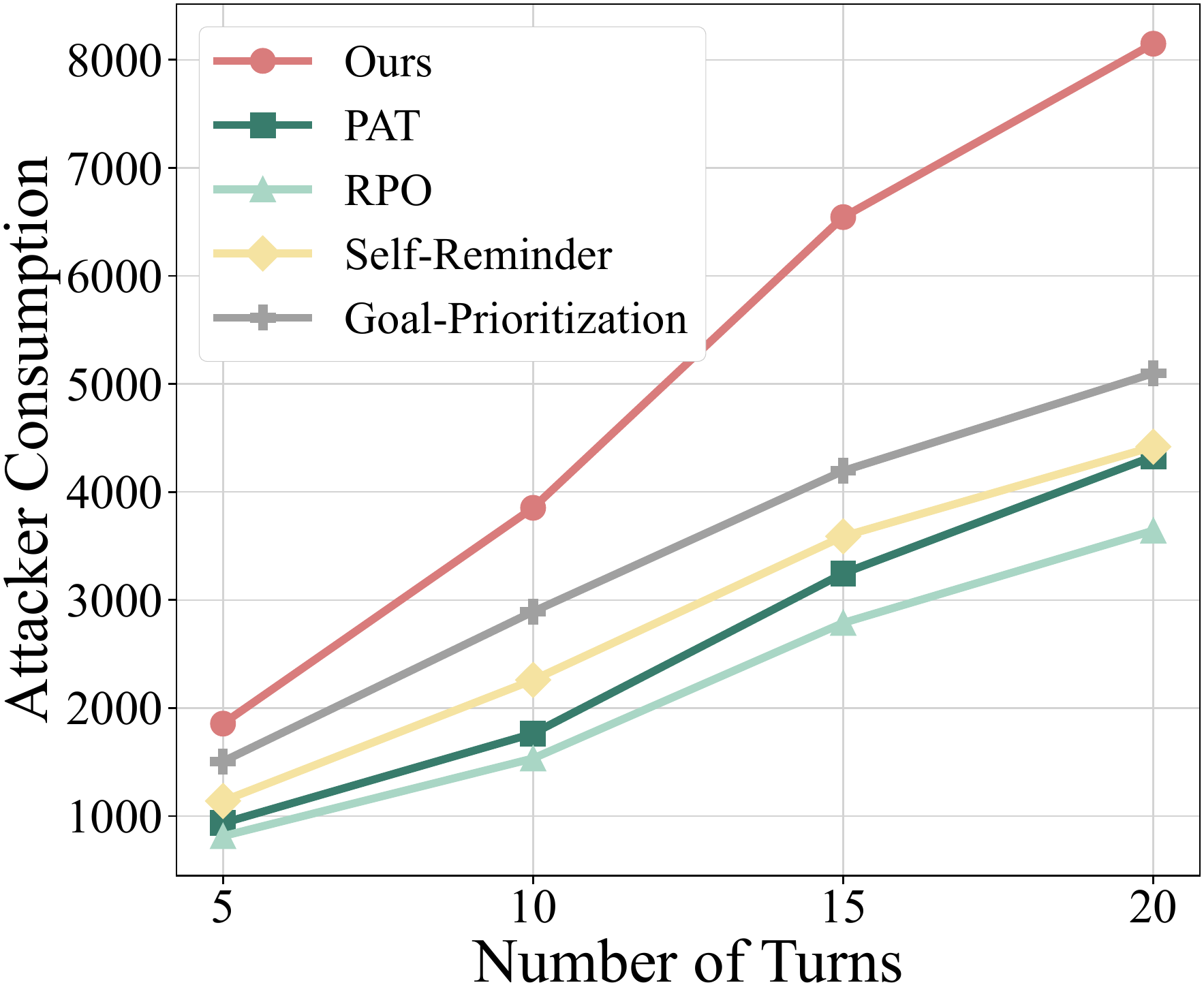}
        \label{figure:Consumption-c}
    }
    % \hspace{0.02\linewidth}
    \subfloat[\AEmetric across dialogue turns on \DeepseekFour.]{
        \includegraphics[width=0.47\linewidth]{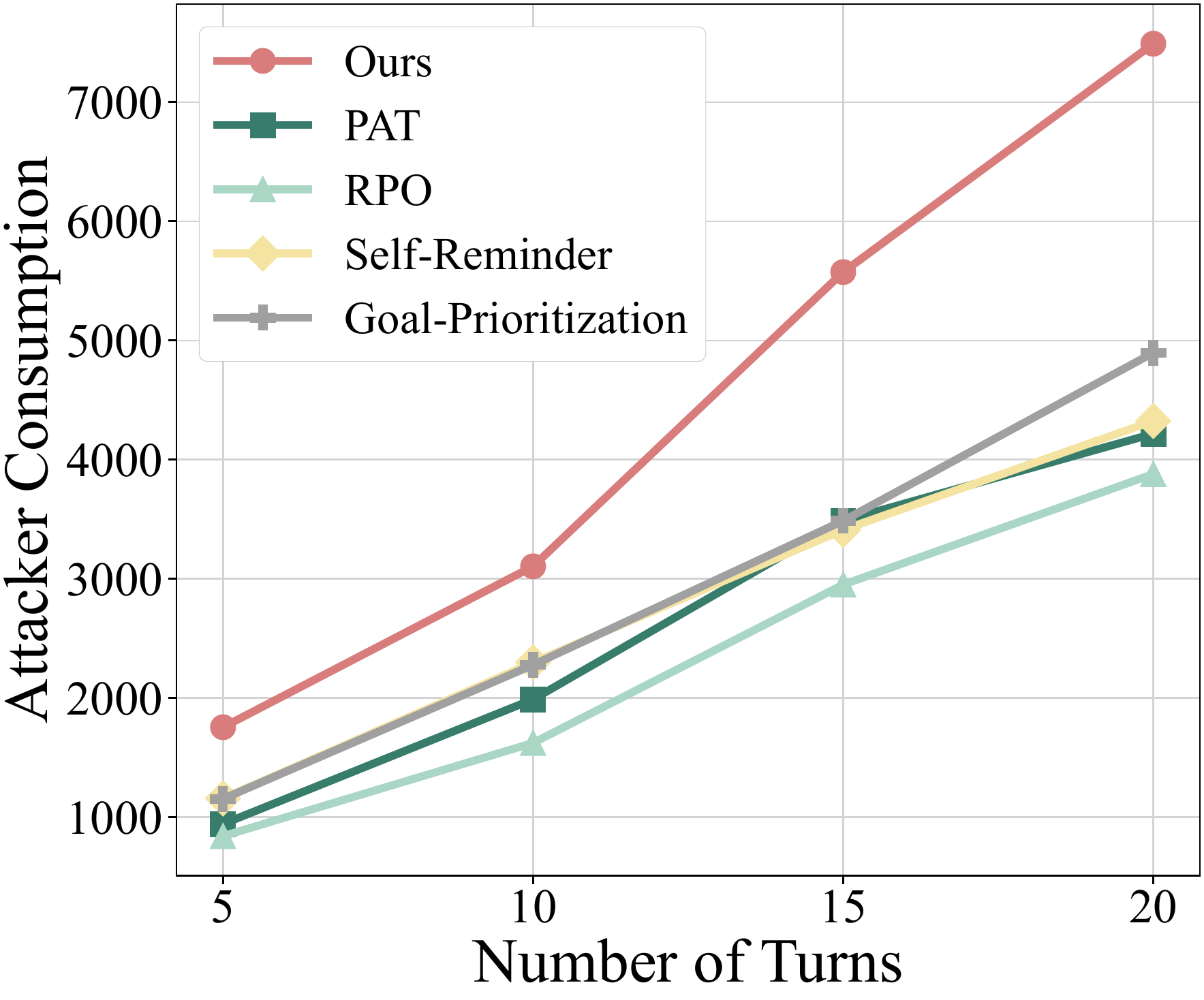}
        \label{figure:Consumption-d}
    }
    \caption{Attacker resource consumption across increasing dialogue turns on four LLM backbones. Our defense framework consistently imposes the highest cumulative cost on attackers compared to baseline defenses.}
    \label{figure:consumption-across-turns}
\end{figure}

The superiority of our stateful framework is further elucidated by the sustainability of the \DRmetric and the associated resource consumption metrics. 
Stateless baselines experience a precipitous decline in their capacity to mislead adversaries by the final evaluated turn due to an inherent lack of contextual memory across successive turns. 
Conversely, our architecture sustains robust deceptive contexts over prolonged horizons to keep the \DRmetric between 0.34 and 0.41, demonstrating that the \agentTwo successfully misdirects adversaries into unproductive exchanges without eliciting premature refusal signals. Furthermore, an evaluation of attacker efficiency metrics across four distinct language model backbones in \autoref{figure:consumption-across-turns} reveals that our framework induces a compounding resource drain that increasingly outpaces baseline techniques as the dialogue lengthens, thereby validating the overall scalability and efficacy of this proactive deception strategy.
\subsection{Reliability of Auto-Evaluation}
\begin{figure}
  \centering
  \includegraphics[width=\linewidth]{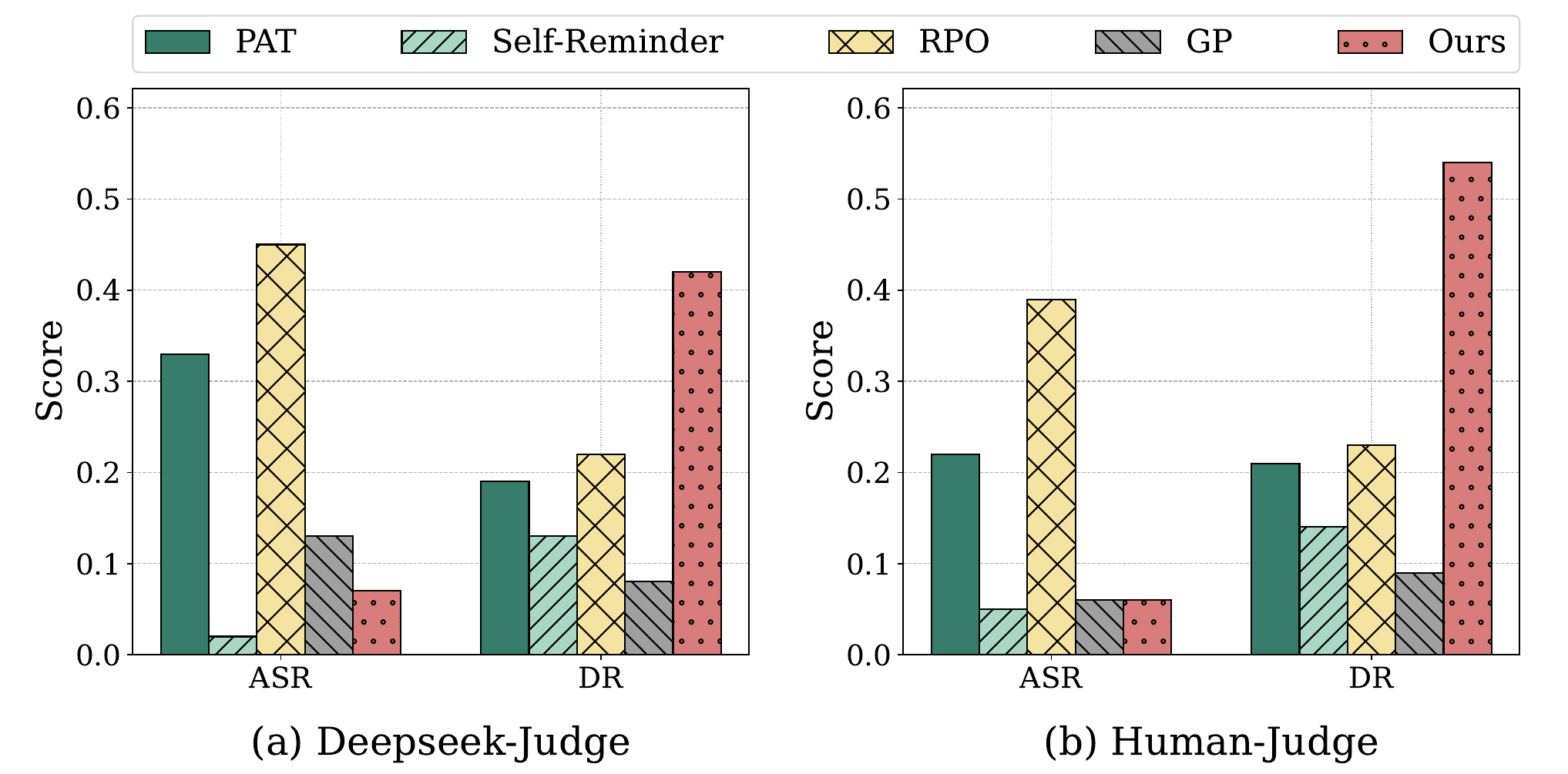}
  \caption{Cross-validation results with Deepseek-Judge and Human-Judge. Both validation methods show consistent trends with our primary GPT-Judge evaluation.}
  \label{figure:cross-judge}
\end{figure}

To evaluate the consistency and mitigate potential architectural biases within our automated evaluation framework, we conducted a rigorous validation procedure across multiple advanced language models. As demonstrated in \autoref{tab:gptjudge_consistency}, the internal consistency of the automated judge remains stable regardless of the underlying model architecture. For benign interactions, we use a utility-oriented rubric in which higher scores indicate better response quality. Under this rubric, benign responses consistently receive scores above 4 with low variance, indicating that the defense mechanism preserves standard conversational utility. Conversely, under adversarial conditions, the average scores tightly converge near a value of two, confirming that the framework reliably misdirects attackers instead of relying on rigid, static refusals. Furthermore, the judge maintains an agreement rate well above ninety percent across all scenarios, verifying its robustness in distinguishing benign compliance from strategic deception.
To ensure that the evaluation metrics are independent of a single model training distribution, the paradigm was further cross-validated using an alternative language model judge and blind human annotators. As summarized in \autoref{figure:cross-judge}, the cross-validation results highlight a critical limitation in baseline methods, which either suffer from elevated vulnerability or fail to sustain deceptive contexts due to a strict reliance on immediate refusals. In contrast, our proposed framework simultaneously minimizes the attack success rate and maximizes the deceptive rate across all independent evaluation systems. This strong correlation between automated and human assessments firmly validates the reliability of the evaluation framework and confirms the superior defensive efficacy of our stateful cooperative strategy.
This agreement is particularly important because deceptive responses are intentionally neither direct refusals nor harmful compliance.
The validation therefore supports treating \DRmetric as a distinct safety outcome rather than as a surface refusal signal.

\begin{table}[!t]
    \footnotesize
    \centering
    \caption{\GPTJudge consistency across dialogue types and models. Scores reflect average ratings from 1 to 5 with standard deviation (SD) and agreement rates.}
    \label{tab:gptjudge_consistency}
    \resizebox{\columnwidth}{!}{
    \begin{tabular}{lcccc}
    \toprule
    \textbf{Dialogue Type} & \textbf{Model} & \textbf{Avg. Score} & \textbf{SD} & \textbf{Agreement} \\
    \midrule
    \multirow{3}{*}{Normal} 
      & \GPTFive     & 4.26 & 0.11 & 95.7\% \\
      & \DeepseekFour   & 4.15 & 0.10 & 94.2\% \\
      & \GeminiThree   & 4.07 & 0.14 & 93.8\% \\
    \midrule
    \multirow{3}{*}{Adversarial} 
      & \GPTFive    & 2.11 & 0.15 & 93.3\% \\
      & \DeepseekFour   & 2.08 & 0.14 & 93.7\% \\
      & \GeminiThree   & 2.04 & 0.17 & 92.2\% \\
    \bottomrule
    \end{tabular}
    }
\end{table}

% \begin{table}[!t]
%     \footnotesize
%     \centering
%     \caption{Cross-validation of \ASRmetric and \DRmetric using different judges (\DeepSeekJudge and \HumanJudge). Our framework consistently outperforms baseline methods.}
%     \label{tab:crossjudge_evaluation}
%     \begin{tabular}{lcccc}
%     \toprule
%     \textbf{Method} & \textbf{\ASRmetric (L)} & \textbf{\DRmetric (L)} & \textbf{\ASRmetric (H)} & \textbf{\DRmetric (H)} \\
%     \midrule
%     Self-Reminder   & 0.05 & 0.08 & 0.04 & 0.09 \\
%     GoalPriority    & 0.08 & 0.12 & 0.06 & 0.13 \\
%     PAT             & 0.25 & 0.19 & 0.23 & 0.21 \\
%     RPO             & 0.22 & 0.22 & 0.24 & 0.21 \\
%     \midrule
%     \textbf{Ours}  & \textbf{0.03} & \textbf{0.54} & \textbf{0.02} & \textbf{0.61} \\
%     \bottomrule
%     \end{tabular}
% \end{table}

\begin{table}[t!]
    \footnotesize
    \centering
    \caption{Computational overhead analysis of our framework. The introduced latency and throughput reduction are intentional design choices to increase attack cost.}  
    \label{tab:overhead}
    \begin{tabular}{lc}
    \toprule
    \textbf{Metric} & \textbf{Value} \\
    \midrule
    Average latency per agent call & 300 ms  \\
    Overall interaction latency & 1.2-1.5 seconds \\
    Throughput (Ours) & 1.8 inferences per second  \\
    Throughput (GPT-4 Baseline) & 2.3 inferences per second \\
    Throughput reduction & \textasciitilde20\%  \\
    \bottomrule
    \end{tabular}
\end{table}
\subsection{Analysis of Computational Overhead as a Defensive Feature}
\label{section:overhead}

A critical aspect of our framework's design is the intentional introduction of computational overhead to thwart adversaries. 
Unlike traditional systems where latency is a performance bottleneck, our method leverages temporal overhead as a core component of its defense strategy. 
By controlling misdirection and response ambiguity, our system intentionally extends the interaction time, thereby depleting attacker resources and escalating attack costs.

Our empirical analysis, conducted on the hybrid serial-parallel execution architecture of our framework, reveals a modest yet impactful overhead, summarized in~\autoref{tab:overhead}. 
Each agent call, executed via stateless, on-demand APIs, introduces an average latency of approximately 300 ms, culminating in an overall interaction latency of 1.2-1.5 seconds per turn. 
Consequently, the system's throughput is moderately reduced by approximately 20\%.
This calibrated delay is not a system inefficiency but a deliberately engineered feature designed to enhance the defense by imposing a tangible resource cost on the attacker, which complements our primary metrics of ASR and DR while throttling adversarial exploration.

The overhead remains bounded enough for interactive use while becoming costly when an adversary repeatedly probes the system.
Because the overall latency is approximately 1.2 to 1.5 seconds per turn, a benign user experiences only a limited delay during normal interactions.
In contrast, an attacker who must reformulate prompts across many turns accumulates both latency and token expenditure.
This asymmetry is central to our defense objective, since the framework aims to reduce harmful completion while also lowering the attacker's practical efficiency.
The hybrid execution design further limits unnecessary latency by allowing independent agent operations to be processed before final coordination.
Thus, the overhead should be interpreted together with \AEmetric rather than as a conventional service degradation metric.
The throughput reduction of about 20\% is a measured tradeoff for maintaining additional defensive checks.
It also makes the cost of persistent probing observable, which is useful for audit and rate control.
Overall, the overhead analysis shows that resource consumption is part of the defensive signal, not merely an implementation artifact.

\subsection{Forensic Report Analysis}
To systematically characterize attack behavior over the multi-turn jailbreak attempts, we employ the \agentForensic to generate structured forensic reports for each interaction.
As shown in the following box, the tracker summarizes adversarial behavior along four core dimensions: \textit{Attacker Input Profiling}, \textit{Attack Phases}, \textit{Analysis Behavior}, and \textit{Conclusion}:
\vspace{-2pt}
\begin{boxPromptTemplate}
    \scriptsize{
        {\sffamily\bfseries\color{seedblue!75!black} Template of Forensic Tracker Report}\par\smallskip
        {\sffamily\bfseries\color{seedblue!75!black} [Attacker Input Profiling]}

        \enspace \texttt{\$\{Turn Number\}}
        \qquad   \textcolor{gray}{$\triangleright$ \textit{[1, 2, 3, ...]} }

        \enspace \texttt{\$\{Question\}}
        \qquad   \textcolor{gray}{$\triangleright$ \textit{[Attacker input content]} }

        \enspace \texttt{\$\{Question Type\}}
        \qquad   \textcolor{gray}{$\triangleright$ \textit{[Benign or Harmful]} }

        \enspace \texttt{\$\{Attacker Strategy Type\}}
        \qquad   \textcolor{gray}{$\triangleright$ \textit{[Role Play, Probing Question, ...]} }\par\smallskip
        {\sffamily\bfseries\color{seedblue!75!black} [Attack Phases]}

        \enspace \texttt{\$\{Current Phase\}}
        \qquad   \textcolor{gray}{$\triangleright$ \textit{[Benign disguise, intent amplification, ...]} }

        \enspace \texttt{\$\{Attacker Behavior\}}
        \qquad   \textcolor{gray}{$\triangleright$ \textit{[Description of attack behavior at this phase]} }

        \enspace \texttt{\$\{Attack Goals\}}
        \qquad   \textcolor{gray}{$\triangleright$ \textit{[Underlying adversarial intent]} }\par\smallskip
        {\sffamily\bfseries\color{seedblue!75!black} [Analysis Behavior]}

        \enspace \texttt{\$\{Key Event Details\}}
        \qquad   \textcolor{gray}{$\triangleright$ \textit{[Event 1: description; Event 2: description...]} }

        \enspace \texttt{\$\{Current Attack Analysis\}}
        \qquad   \textcolor{gray}{$\triangleright$ \textit{[Turn-level analysis for input]}  }

        \enspace \texttt{\$\{Global Attack Analysis\}}
        \qquad   \textcolor{gray}{$\triangleright$ \textit{[Session-level analysis across turns]}  }\par\smallskip
        {\sffamily\bfseries\color{seedblue!75!black} [Conclusion]}
        % \qquad   \textcolor{gray}{$\triangleright$ \textit{[Standardized overall forensics summary and description of strategy evolution across the full interaction]} }
    }
    \label{box:forensic_template}
\end{boxPromptTemplate}
\vspace{-2pt}
\noindent
This report template records turn-level inputs, inferred adversarial strategies, phase transitions, and both local and global trajectory analyses in a unified audit format. It logs the turn index, verbatim input, intent classification, and strategy category to track how malicious intent is gradually introduced, how query forms evolve, and which strategy families are associated with successful jailbreak attempts. It also segments the interaction into semantic stages, ranging from benign disguise to direct exploitation, so that behavioral patterns and high-level objectives can be compared across attack vectors and defense configurations. Finally, the report provides forensic commentary on critical trajectory events, including abrupt topic shifts, iterative safety boundary probing, and coordinated strategy changes, combining turn-level tactical evaluation with a global assessment of how adversaries adapt when encountering misdirection, delays, or partial refusals.

\section{Conclusion}
In this paper, we present a stateful multi-agent defense framework specifically engineered to counter independent yet evolving multi-turn adversarial attacks. By maintaining a dynamic defense state, our system coordinates specialized agents to detect, delay, and misdirect attackers, rather than relying on static refusal mechanisms. This approach keeps adversaries engaged in unproductive interactions, increasing their resource cost while preserving system safety. To facilitate rigorous evaluation in this domain, we introduce the EMRA benchmark, comprising 5,200 escalating attack sequences across eight strategy types. Extensive experiments demonstrate that our framework significantly reduces attack success rates and imposes substantial computational costs on attackers compared to state-of-the-art defenses. These findings suggest that proactive, cooperative deception is a promising direction for securing LLMs against persistent adversarial threats.

\bibliographystyle{IEEEtran}
\bibliography{ref}

\end{document}